\documentclass[10pt]{article}

\usepackage{amsmath}

\usepackage{array}

\usepackage{appendix}

\usepackage{tocloft}                   

\usepackage{graphicx}

\usepackage{amsfonts}

\usepackage{amssymb}

\usepackage{mathrsfs}

\usepackage{yfonts}

\usepackage{euscript}

\usepackage{centernot}                 

\usepackage{ifsym}                     

\usepackage{lmodern}                   

\usepackage{upgreek}

\usepackage{mathtools}

\usepackage{bm}

\usepackage{undertilde}

\usepackage{dutchcal}

\usepackage{amsfonts}

\usepackage{amssymb}

\usepackage{mathrsfs}

\usepackage{upgreek}

\usepackage{mathtools}

\usepackage{color}

\usepackage{slantsc}
\usepackage{calligra}

\usepackage{bbold}          

\usepackage[T1]{fontenc}

\usepackage{epsf}

\usepackage{latexsym}

\usepackage{tipa}

\usepackage{makeidx}

\makeindex

\def\b{\begin{equation}}

\def\e{\begin{equation}}

\def\be{\begin{equation}}              

\def\ee{\end{equation}}

\def\beq{\begin{equation}}

\def\eeq{\end{equation}}

\def\bea{\begin{eqnarray}}

\def\eea{\end{eqnarray}}

\def\half{\mbox{$\frac{1}{2}$}}

\def\m{\mbox{ }}

\def\mma {\m , \m \m }

\def\!{\hspace{-1.6667em}}

\def\Leftrightarrows{\m \m \Leftrightarrow \m \m}

\def\n{\noindent}

\def\u{\underline}

\def\uc{\underbracket}   
\def\uo{\utilde}         

\def\w{\widetilde}

\def\s{\stackrel}

\def\bic{\mbox{\boldmath$c$}}

\def\biA{\mbox{\boldmath$A$}}

\def\biB{\mbox{\boldmath$B$}}   

\def\biC{\mbox{\boldmath$C$}}              

\def\biD{\mbox{\boldmath$D$}}

\def\biE{\mbox{\boldmath$E$}}

\def\sbiF{\mbox{\boldmath \scriptsize $F$}} 

\def\biG{\mbox{\boldmath  $G$}}             %
\def\biF{\mbox{\boldmath $F$}}             %

\def\biH{\mbox{\boldmath$H$}}

\def\biI{\mbox{\boldmath$I$}}

\def\biJ{\mbox{\boldmath$J$}}

\def\biK{\mbox{\boldmath$K$}}

\def\biL{\mbox{\boldmath$L$}}

\def\biM{\mbox{\boldmath$M$}}

\def\biN{\mbox{\boldmath$N$}}

\def\biO{\mbox{\boldmath$O$}}

\def\biW{\mbox{\boldmath$W$}}

\def\biX{\mbox{\boldmath$X$}}

\def\biY{\mbox{\boldmath$Y$}}

\def\biZ{\mbox{\boldmath$Z$}}

\def\sbiU{\mbox{\ttfamily\fontseries{b}\selectfont U}}                   %

\def\sbiO{\mbox{\ttfamily\fontseries{b}\selectfont O}}

\def\bgamma{\mbox{\boldmath$\gamma$}}               
\def\bTheta{\mbox{\boldmath$\Theta$}} 

\def\bxi{\mbox{\boldmath$\xi$}}

\def\bphi{\mbox{\boldmath$\phi$}} 

\def\balpha{\mbox{\boldmath$\alpha$}} 

\def\sbalpha{\mbox{\scriptsize\boldmath$\alpha$}} 

\def\mA{\mbox{A}}  

\def\mB{\mbox{B}}  

\def\mH{\mbox{H}} 

\def\mL{\mbox{L}}

\def\mS{\mbox{S}}                        

\def\mT{\mbox{T}} 

\def\mV{\mbox{V}}

\def\mZ{\mbox{Z}}

\def\me{\mbox{e}}

\def\mm{\mbox{m}}

\def\mp{\mbox{p}}

\def\mx{\mbox{x}}

\def\bE{\mbox{\bf E}}

\def\bG{\mbox{\bf G}}                     

\def\bT{\mbox{\bf T}}

\def\bV{\mbox{\bf V}}

\def\bX{\mbox{\bf X}}

\def\bg{\mbox{\bf g}}

\def\bm{\mbox{\bf m}}

\def\bx{\mbox{{\bf x}}}

\def\bsigma{\mbox{\boldmath$\sigma$}}                   %

\def\bupPhi{\mbox{\boldmath$\Phi$}}                     

\def\bupxi{\mbox{\boldmath$\xi$}}                       

\def\sbupxi{\mbox{\scriptsize\boldmath$\xi$}}                       

\def\bupXi{\mbox{\boldmath$\Xi$}}  

\def\bupPhi{\mbox{\boldmath$\Phi$}}  

\def\bupOmega{\mbox{\boldmath$\Omega$}}                   

\def\scg{\mbox{\boldmath ${\cal g}$}}

\def\sch{\mbox{\boldmath ${\cal h}$}}

\def\sck{\mbox{\boldmath ${\cal k}$}}

\def\fe{\mbox{\sffamily e}}

\def\ff{\mbox{\sffamily f}}

\def\fii{\mbox{\sffamily i}}

\def\fA{\mbox{\sffamily A}}           

\def\bfS{\mbox{\bf\sffamily S}}

\def\bfB{\mbox{\bf\sffamily B}}

\def\sbfB{\mbox{\scriptsize\bf\sffamily B}}

\def\fC{\mbox{\sffamily C}}

\def\fD{\mbox{\sffamily D}}

\def\fG{\mbox{\sffamily G}}

\def\fO{\mbox{\sffamily O}}

\def\cC{{\mathscr C}}

\def\cD{{\mathscr D}}

\def\cE{{\mathscr E}}

\def\cH{{\mathscr H}}

\def\cK{{\mathscr K}}

\def\cL{{\mathscr L}}

\def\cP{{\mathscr P}}

\def\cR{{\mathscr R}}

\def\cS{{\mathscr S}}

\def\sd{\mbox{\scriptsize d}}

\def\se{\mbox{\scriptsize e}}

\def\sf{\mbox{\scriptsize f}}

\def\si{\mbox{\scriptsize i}}

\def\sn{\mbox{\scriptsize n}}

\def\sp{\mbox{\scriptsize p}}

\def\su{\mbox{\scriptsize u}}

\def\sv{\mbox{\scriptsize v}}

\def\sw{\mbox{\scriptsize w}}

\def\sx{\mbox{\scriptsize x}}

\def\sD{\mbox{\scriptsize D}}

\def\sS{\mbox{\scriptsize S}}

\def\sT{\mbox{\scriptsize T}}

\def\sfk{\mbox{\sffamily{\scriptsize k}}}     

\def\sfA{\mbox{\sffamily{\scriptsize A}}}     

\def\sfC{\mbox{\sffamily{\scriptsize C}}}     

\def\sfD{\mbox{\sffamily{\scriptsize D}}}      

\def\sfG{\mbox{\sffamily{\scriptsize G}}}      

\def\sfK{\mbox{\sffamily{\scriptsize K}}}      

\def\sfO{\mbox{\sffamily{\scriptsize O}}}      

\def\sbX{\mbox{{\bf \scriptsize X}}}

\def\bfO{\mbox{{\bf \sffamily O}}}                                    
\def\bfQ{\mbox{{\bf \sffamily Q}}}                                    

\def\bfP{\mbox{{\bf \sffamily P}}}                                    

\def\bfa{\mbox{{\bf \sffamily a}}}                                    

\def\bscF{\mbox{{\boldmath \scriptsize${\cal F}$}}}                               

\def\bscS{\mbox{\boldmath \scriptsize${\cal S}$}}                               

\def\bscf{\mbox{{\boldmath ${\cal f}$}}}                               

\def\bsce{\mbox{{\boldmath ${\cal e}$}}}                               

\def\bscs{\mbox{{\boldmath ${\cal s}$}}}                               

\def\bsci{\mbox{{\boldmath ${\cal i}$}}}                               

\def\ttO{\mbox{\tt O}}

\def\Thomas{\,\,\mbox{\textcircled{$\rightarrow$}}\,\,}

\def\TwoWay{\,\,\mbox{\textcircled{$\leftrightarrow$}}\,\,}

\def\sumi2{\sum\mbox{}_{\mbox{}_{\mbox{\scriptsize $i$=1}}}^2}

\def\sumi3{\sum\mbox{}_{\mbox{}_{\mbox{\scriptsize $i$=1}}}^3}

\def\sumABcycles3{\sum\mbox{}_{\mbox{}_{\mbox{\scriptsize cycles $A,B$=1}}}^{3}}

\def\sumCDcycles3{\sum\mbox{}_{\mbox{}_{\mbox{\scriptsize cycles $C,D$=1}}}^{3}}

\def\sumj3{\sum\mbox{}_{\mbox{}_{\mbox{\scriptsize $j$=1}}}^3}

\def\sumk3{\sum\mbox{}_{\mbox{}_{\mbox{\scriptsize $k$=1}}}^3}

\def\prodiA1{\prod\mbox{}_{\mbox{}_{\mbox{\scriptsize $i$=1}}}^{A - 1}}

\def\bigtimes{\mbox{\Large $\times$}}

\def\d{\textrm{d}}                                                  

\def\pa{\partial}                                                   

\def\es{\m = \m}

\def\:={\m := \m}

\def\=:{\m =: \m}

\def\peq{\m \mbox{`='} \m}

\def\speq{\m \peq \m}

\def\FrI{\mbox{$\mathfrak{I}$}}                                

\def\lFrs{\mathfrak{S}}                                        

\def\FrT{\mathfrak{T}}                                         

\def\FrC{\mbox{$\mathfrak{C}$}}                                
                                                       
\def\FrX{\mathfrak{X}}                                         
\def\FrY{\mathfrak{Y}}                                         

\def\FrS{\mbox{\Large $\mathfrak{s}$}}                         
\def\FrU{\mbox{$\mathfrak{U}$}}                                
\def\bFrV{\mbox{\boldmath$\mathfrak{V}$}} 

\def\FrV{\mbox{$\mathfrak{V}$}}                                
\def\FrW{\mbox{$\mathfrak{W}$}}                                
\def\FrD{\mbox{$\mathfrak{D}$}}	                               
 
\def\Frm{\mbox{\Large $\mathfrak{m}$}}                         
\def\FrN{\mbox{$\mathfrak{N}$}}                                
\def\lFrg{\mbox{\Large$\mathfrak{g}$}}                         
\def\FrK{\mathfrak{K}}                                         
\def\FrN{\mathfrak{N}}                                         
                                                  
\def\Frg{\mbox{\normalsize $\mathfrak{g}$}}                    

\def\Frk{\mbox{\scriptsize $\mathfrak{K}$}}                    

\def\Frh{\mbox{$\mathfrak{h}$}}                                

\def\FrF{\mbox{\boldmath$\mathfrak{F}$}}                       

\def\FrT{\mbox{\boldmath$\mathfrak{T}$}}                       
\def\Hilb{\mbox{{\boldmath$\mathfrak{H}$}ilb}}                 
\def\Frc{\mbox{\Large $\mathfrak{c}$}}                         

\def\scC{\mbox{\scriptsize $\cC$}}                    
\def\bscS{\mbox{\boldmath\scriptsize ${\cal S}$}}

\def\Obs{\FrO\mbox{bs}}                             %

\def\UnresObs{\FrU\mbox{nres-}\FrO\mbox{bs}}                                 %

\def\FullObs{\FrF\mbox{ull-}\FrO\mbox{bs}}                                  %

\def\DiracObs{\FrD\mbox{irac-}\FrO\mbox{bs}}                                  %

\def\Sec{\bscS\mbox{\bf e}}                                

\def\FrQ{\mbox{\Large $\mathfrak{q}$}}                               
\def\bFrF{\mbox{\boldmath$\mathfrak{F}$}}                            %

\def\bFrG{\mbox{\boldmath$\mathfrak{G}$}}                            %
												  
\def\Phase{\mbox{{\boldmath$\mathfrak{P}$}hase}}                     

\def\bFrR{\mbox{\boldmath$\mathfrak{R}$}}                            
\def\Rig-Phase{\bFrR\mbox{ig-}\Phase}                                
 																	 
\def\Spacetime{\FrS\mbox{pacetime}}                                  

\def\bFrM{\mbox{\boldmath${\mathfrak{M}}$}}

\def\bFrR{\mbox{\boldmath$\mathfrak{R}$}}                            

\def\bFrR{\mbox{\boldmath$\mathfrak{R}$}}                            

\def\1mat{\u{\u{1}}}                                                 

\def\Positive-Modespace{\mbox{{\boldmath$\mathfrak{M}$}odespace$^+$}}

\def\POSITIVE-MODESPACE{\mbox{{\boldmath$\mathfrak{M}$}ODESPACE$^+$}}
                                                                                                                             														
\def\bFrS{\mbox{\Large $\mathfrak{s}$}}                              
			
\def\bFrG{\mbox{\boldmath $\mathfrak{G}$}}                                    %

\def\FrO{\mbox{$\mathfrak{O}$}}                                      

\def\bFrO{\mbox{\boldmath$\mathfrak{O}$}}                            

\def\lattice{\mbox{\bf\Large$\mathfrak{L}$}}                                      

\def\Kin-Hilb{\mbox{{\boldmath$\mathfrak{K}$}in-\Hilb}}                     

\def\Mid-Hilb{\mbox{{\boldmath$\mathfrak{M}$}id-\Hilb}}                     

\def\Dyn-Hilb{\mbox{{\boldmath$\mathfrak{D}$}yn-\Hilb}}                     

\def\5Star{\mbox{\Large$\star$}}              

\begin{document}

\begin{center}

\Huge{\bf A LOCAL RESOLUTION OF}

\vspace{.1in}

\normalsize

\Huge{\bf THE PROBLEM OF TIME}

\vspace{.1in}

\Large{\bf XIV. Grounding on Lie's Mathematics}

\vspace{.1in}

{\large \bf E.  Anderson} 

\end{center}

\begin{abstract}

In a major advance and simplification of this field, we show that A Local Resolution of the Problem of Time  
                                   -- which can also be viewed as A Local Theory of Background Independence --  
can at the classical level be described solely in terms of Lie's Mathematics.  									  
This comprises 
i) Lie derivatives to encode Relationalism, including via solving the generalized Killing equation. 
ii) Lie brackets to formulate Closure, via Lie's Algorithm suitably extended to accommodate insights of Dirac and from Topology,     
    producing generator Lie algebraic structures: Lie algebras or Lie algebroids.   
iii) Observables defined by Lie brackets relations, 
which can be recast as explicit PDE systems to be solved using the Flow Method, 
and constitute observables Lie algebras. 
Lattices of constraint algebraic substructures furthermore induce dual lattices of observables subalgebras.  
iv) The `passing families of theories through the Dirac Algorithm' approach to Spacetime Construction from Space, 
and to obtaining more structure from less each internally to each of space and spacetime separately, 
are identified as deformations that work selectively when Lie Algebraic Rigidity is encountered. 
v) Reallocation of Intermediary-Object (RIO) Invariance: the general Lie Theory's commuting-pentagon analogue of posing Refoliation Invariance for GR. 
i) to v) cover respectively the Relationalism, Closure, Observables, Deformations and RIO super-aspects of Background Independence, 
Lie Theory moreover already collates i) to iii) and the internal case of iv) as multiple interacting aspects. 
The Problem of Time's multiple interacting facets are then explained as, firstly, 
the result of having two copies of this Lie collation, one for each of the spacetime and `space, dynamics or canonical' primalities.
Secondly, a Wheelerian two-way route between these two primalities, comprising v) and the `spacetime from space' version of iv). 
We further develop the Comparative Theory of Background Independence in this manner. 
We can even give a `pillars of the Foundations of Geometry' parallel of our Background Independence super-aspects, 
including both new and well-established foundational pillars.  

\end{abstract}

$^1$ dr.e.anderson.maths.physics *at* protonmail.com

\section{Introduction}\label{Introduction}

\subsection{What we mean by Lie's Mathematics}

The enclosed account of Lie's Mathematics  \cite{Lie, Yano55, Jacobson, Serre-Lie, NR66, Yano70, CM, M08, Lee2, BCHall}
suffices to construct A Local Resolution \cite{ALett, ABook, I, II, III, IV, V, VI, VII, VIII, IX, X, XII, XIII} 
of the Problem of Time \cite{Battelle, DeWitt67, DiracObs, Dirac51, Dirac58, Dirac, K92, I93, APoT, APoT2, ABook} (ALRoPoT), 
which in turn can be reformulated as \cite{APoT3, ABook, A-Killing, A-Cpct, A-CBI, I, I, II, III, IV, V, VI, VII, VIII, IX, X, XII, XIII} 
A Local Theory of Background Independence \cite{A64, A67, Giu09} (ALToBI) at the classical level.   
This represents a major advance and simplification of the Problem of Time and Background Independence field of study.  
The following Lie structures are employed in this venture.

\m 

\n i) {\bf Lie derivatives} \cite{Pauli, Sleb, Dantzig, Yano55, Yano70} are used to encode Relationalism, 
with solving the {\bf generalized Killing equation} \cite{Yano55, Yano70} of further use for encoding Configurational or Spacetime Relationalism.

\m 

\n ii) {\bf Lie brackets} are introduced to formulate Closure, and assess whether this is attained using {\bf Lie's Algorithm} \cite{Lie} 
suitably extended to accommodate Dirac \cite{Dirac, HTBook} and topological \cite{Bertlmann, ABook} insights.
If Closure is attained, the end product is a {\bf generator Lie algebraic structure}. 
This means either a {\bf Lie algebra}   \cite{Jacobson, Serre-Lie, FHBook, BCHall} 
               or a {\bf Lie algebroid} \cite{CM},   
such as the Dirac algebroid \cite{Dirac51, Dirac58, Dirac} formed by GR's constraints.  

\m 

\n iii) {\bf Observables} \cite{ABook} are, given a state space $\FrS$, suitably smooth functions thereover. 
In the presence of a Lie group $\lFrg$ of transformations `to be held to be irrelevant to the modelling' acting on $\FrS$, moreover, 
furtherly useful notions of observables \cite{DiracObs, K93, ABook, III, VIII, X} are to have {\bf zero-commutant Lie brackets with the generators} of $\lFrg$. 
These Lie brackets relations can furthermore be recast as \cite{AObs2, PE-1, DO-1, VIII} as explicit first-order linear PDE systems. 
The {\bf Flow Method} \cite{Lie, G63, John, Olver2, M08, Lee2, Olver} can be evoked to solve these; 
interpretations for flows include {\bf congruences of integral curves} and {\bf 1-parameter subgroups} of Lie groups. 
This particular application of the Flow Method essentially constitutes {\bf Lie's Integral Approach to Invariants} \cite{Lie, G63, Olver2, Olver} 
albeit elevated to restricting functional dependence on a function space over the geometry in question, 
and with the physical case being instead over phase space or the space of spacetimes. 
Observables moreover form Lie algebras \cite{PE-1, PE-2-3, DO-1}: {\bf observables Lie algebras}. 

\m 

\n Each theory's  {\bf lattice of constraint algebraic substructures} \cite{AObs3, ABook, DO-1, III} additionally 
induces a {\bf dual lattice of observables          subalgebras}   \cite{AObs3, ABook, DO-1, III}.

\m  

\n iv) The `passing families of theories through the Dirac Algorithm' approach to Spacetime Construction from Space \cite{RWR, AM13, ABook, IX}, 
and to obtaining more structure from less each internally to each of space and spacetime separately, 
are identified as \cite{Higher-Lie} {\bf Lie algebraic structure deformation procedures} \cite{G64, NR66, CM} 
that work selectively when {\bf Lie Algebraic Rigidity} \cite{G64, NR66, CM} is encountered.

\m 

\n v) Reallocation of Intermediary-Object (RIO) Invariance \cite{Higher-Lie} 
is the general {\bf Lie Theory's commuting-pentagon} analogue of posing Refoliation Invariance \cite{T73, I93, ABook} for GR.
This pentagon amounts to whether switching which intermediary object one proceeds via amounts to at most a difference by an automorphism of the final object.  

\m 

\n This is an opportune place at which remind the reader that the five super-aspects of this Series' notion of Background Independence are i)  Relationalism, 
                                                                                                                                           ii)  Closure, 
																									                                       iii) Observables, 
																									                                       iv)  Construction from Deformations via Rigidity, 
																									                                   and v)   RIO, 
by which the above account does indeed cover all five.  

\m 

\n N.B.\ also that all bar algebroids, iv) and v) can be expected of Fresher Graduates in Physics or in Continuum Mathematics. 
Much of the Problem of Time and Background Independence -- an exciting field of study -- 
is hereby now prised open to a very accessible level for the very first time.

\subsection{Outline of this Article}

\n Lie's original mathematical structures are outlined in Sec \ref{Lie-Position}. 
His original modelling assumptions are in Sec \ref{Lie-Modelling}.
On the one hand, we argue to build on Lie's locality modelling assumption; e.g.\ the `local' in ALRoPoT is a larger version of this. 
On the other hand, we also remove Lie's other two modelling assumptions to take into account subsequent developments in each of Topology and contemporary Theoretical Physics. 
Various brief points concerning topological spaces, topological manifolds, Functional Analysis and Relativity are collected here for this purpose.   

\m 

\n Sec \ref{Diff-Manifold} further sets up (sufficiently) smooth differentiable manifolds.  
A first application of this is Sec \ref{Lie-Deriv}'s presentation of the Lie derivative. 
In Sec 8, we use the Lie derivative within a $\lFrg$-act $\lFrg$-all scheme to correct state space objects (Article II's main subject).     
A second application is in Sec 7's outline of Lie group and Lie algebras.  
Sec \ref{GKEs}'s consideration of generalized Killing equations then makes use of all preceding parts of this paragraph. 
These last two considerations produce reduced state spaces, 
motivating consideration of spaces that at least locally admitting differentiable structure, as outlined in Sec \ref{+Diff}. 

\m 

\n Sec \ref{LA} next presents `Lie's Algorithm' for Generator Closure. 
This generalizes to the general Lie bracket setting (see also Article X) 
the more well-known subcase of using the Dirac Algorithm \cite{Dirac, HTBook, ABook} to assess Constraint Closure (as per Articles III and VII).  
Lie algebraic structures arising from the Lie Algorithm are considered in Sec \ref{LAS}, 
including introduction of Lie algebroids and discussion of topological terms. 
Sec \ref{Split} then proceeds to consider split versions of such algebraic structures. 

\m  

\n Sec \ref{EitoO} sets up observables in the general Lie-theoretic setting.  
Sec \ref{Order} outlines order and lattices for use in Lie Theory, applied to both closure and, dually, to observables 
(supporting Article III).    
Observables furthermore admit a further PDE formulation (Sec \ref{LIToI}), to be approached using the Flow Method 
(see Articles VIII and X for details).

\m 

\n Sec \ref{LR} subsequently outlines Lie algebraic structure deformation procedures \cite{G64}, 
and Rigidity for both Lie algebras \cite{G64, NR64} and Lie algebroids \cite{CM}. 
This is next applied to the Lie Algorithm, enabling this to work as a selection principle, at least when Lie Algebraic Rigidity is encountered.
This provides a more general theory for the mathematical phenomena observed in Constructability aspects of Background Independence. 
\n In Sec \ref{RIO}, we finally pose the universal (theory-independent) analogue of GR's Refoliation Invariance (\cite{T73} and Article XII), 
{\it Reallocation of Intermediary-Object Invariance} for general Lie Theory.   

\m 

\n Our concluding Sec \ref{Conclusion} explains how Lie Theory already collates i) to iii) and the internal case of iv) as multiple interacting aspects; 
this collation forms the `Lie 3-star digraph'. 
The Problem of Time's multiple interacting facets \cite{K92, I93, ABook} are then explained as, firstly, 
the result of having two copies of this Lie collation, one for each of the spacetime and `space, dynamics or canonical' primalities.
Secondly, a Wheelerian \cite{Battelle} two-way route between these two primalities, comprising v) and the `spacetime from space' version of iv). 
We further outline the Comparative Theory of Background Independence \cite{A-Killing, A-Cpct, A-CBI} in this vein. 

\m 

\n We can even give a `pillars of (the Foundations of) Geometry' parallel of our Background Independence super-aspects. 
`Pillar of Geometry' is a term used by e.g.\ Stillwell \cite{Stillwell}, 
with the four traditional such being the Euclidean, Klein's Erlangen, Linear-Algebra-based, and Projective pillars. 
We view this as a list that is ever-open to additions, along similar lines to Wheeler's `many routes to relativity' \cite{Battelle, MTW}. 
Indeed, we include both new and well-established foundational pillars as counterparts of the four Background Independence 
super-aspects picked out by the above `Lie 3-star digraph' collation.  

\m 

\n The current Article's more technical content make it a more suitable recipient than Article XIII for pointers 
to subsequent global Problem of Time and Background Independence work. 
Such subsections are however kept brief and `starred off' as not required for self-contained understanding of the current local Series.


\section{Lie's position from a structural point of view}\label{Lie-Position}

\subsection{First-order PDEs}\label{FOP}

\n{\bf Remark 1} Lie's work \cite{Lie} is foremost on differential equations.
He considers in particular first-order PDEs. 
For a single such, 
\be 
\sum_{A = 1}^N a^{A}(x^{B}, \phi) \pa_A \phi = b(x^{B}, \phi)  \m ,
\ee 
Lie makes use of how one can locally straighten with respect to one variable, permitting integration. 
Via e.g.\ Eisenhart \cite{Eisenhart33} and Yano \cite{Yano55}, such a result is termed `useful Lemma' in Stewart \cite{Stewart}.  

\m 

\n A system of $A = 1$ to $M$ such PDEs, 
\be 
\sum_{B = 1}^N a^{AB}(x^{C}, \phi) \pa_B \phi = b^A(x^{C}, \phi)
\ee
can moreover be `straightened out' one equation at a time. 

\m 

\n Four structural developments branch off at this point. 

\m 

\n{\bf Structural development 1} Each such PDE system $\cP$ admits an ODE formulation by Lagrange's Method of Characteristics \cite{Lagrange, CH2, John}. 
This moreover admits a more modern and differential-geometrically valid `flow' reinterpretation \cite{Stewart, Lee2}. 
Via `{\bf local Lie dragging}', integral curve flowlines ensue.
`Local Lie dragging' was moreover subsequently formalized \cite{Pauli, Sleb, Dantzig} as the notion of Lie derivative \cite{Yano55, Yano70} (Sec \ref{Lie-Deriv}).  

\m 

\n{\bf Structural development 2} There is relation between `{\bf infinitesimal transformation groups}' $\lFrg_{\sT}$ 
and such PDE systems $\cP$.

\m 

\n {\bf Forward route} Lie \cite{Lie} uses differentiation and elimination to provide a $\cP$ for each $\lFrg_{\sT}$.  

\m 

\n {\bf Backward route 1} The above integral method sends such $\cP$ back to the $\lFrg_{\sT}$ of solutions for it.  

\m 

\n{\bf Backward route 2} Given a geometry, Killing \cite{Killing} and successors \cite{Yano55, Yano70} 
moreover construct a specific geometrical $\cP$ whose solutions form the corresponding geometrical $\lFrg_{\sT}$.  

\m 

\n{\bf Remark 2} A more modern presentation of `straightening out' involves the {\bf exponential map} \cite{Lee2} (see Sec \ref{Lie-Lie}) 
applied to each of our $\lFrg_{\sT}$'s 1-parameter subgroups (a further interpretation for flows).

\subsection{The Lie bracket}

\n{\bf Structural development 3} The {\bf Lie bracket} 
\be 
\mbox{\bf |[} \,  \m \mbox{\bf ,} \, \m \mbox{\bf ]|}
\ee
arises naturally in two ways (at least).  

\m 

\n i) Its zeroness ensures that {\bf second-order terms vanish for our infinitesimal transformations} (see Sec \ref{IC}).   

\m 

\n ii) It arises in the {\bf integrability condition}:  
\be
\mbox{if \m $X_i$ \m and \m $X_j$ \m solve \m $\cP$ \mma $\mbox{\bf|[} X_i \mbox{\bf ,} \, X_j \mbox{]|}$ \m also solves \m $\cP$}  \m .  
\ee

\subsection{Some diversities among groups and algebras}

\n{\bf Structural development 4} Lie makes distinction between continuous versus discrete transformations, 
aiming to concentrate on the {\bf continuous transformations}.  

\m 

\n As regards groups which are a mixture of continuous nor discrete, he has a notion of what we now call {\bf connected components}: 
(path-)connected to the identity, by which these reduce to a connected component that is a purely-continuous group.  
%
%
Nontrivial {\bf Lie groups} $\lFrg$ over $\mathbb{R}$ or $\mathbb{C}$ can be continuous or a mixture.

\m 

\n Finally, looking at an infinitesimal neighbourhood of the identity of a Lie group $\lFrg$ gives the corresponding {\bf Lie algebra} $\Frg$: 
the modern version of `infinitesimal transformation group' $\lFrg_{\sT}$.     
(This `looking' focuses on the connected component and this giving is via use of the exponential map.)     
Lie algebras and Lie groups have subsequently become a large field of study; see \cite{Stillwell-Lie} if in need of an undergraduate-level introduction, 
\cite{Gilmore} for a graduate-school Physics text, or \cite{Jacobson, Serre-Lie, Cartan55, FHBook, BCHall} for more advanced texts.  


\section{Lie's modelling assumptions versus modern context}\label{Lie-Modelling}

\subsection{Lie's modelling assumptions}\label{Lie-Modelling-1}

Lie \cite{Lie} worked with the following three modelling assumptions. 

\m 

\n{\bf Modelling Assumption 1)} {\bf Free generic relocalization}, meaning that one is willing to lose part of one's original domain or `neighbourhood' to go through with one's proof. 
This technique is moreover amenable to multiple successive applications within a single given proof. 
By this, domains and `neighbourhoods' are often small; 
these are moreover always taken to be what we would now, in the next subsection's more modern terms, call (path-)connected.   

\m

\n{\bf Modelling Assumption 2)}  {\bf Analyticity} of the geometrical objects under study.  
I.e.\ $\FrC^{\omega}$                functions are in use in the real    case, 
or    $\FrO^{\omega}$ -- holomorphic functions --  in        the complex case.

\m 

\n{\bf Modelling Assumption 3)} {\bf Domains and `neighbourhoods' are used without naming them}.  This is an efficient simplifier of workings. 
%
{\begin{figure}[!ht]
\centering
\includegraphics[width=0.5\textwidth]{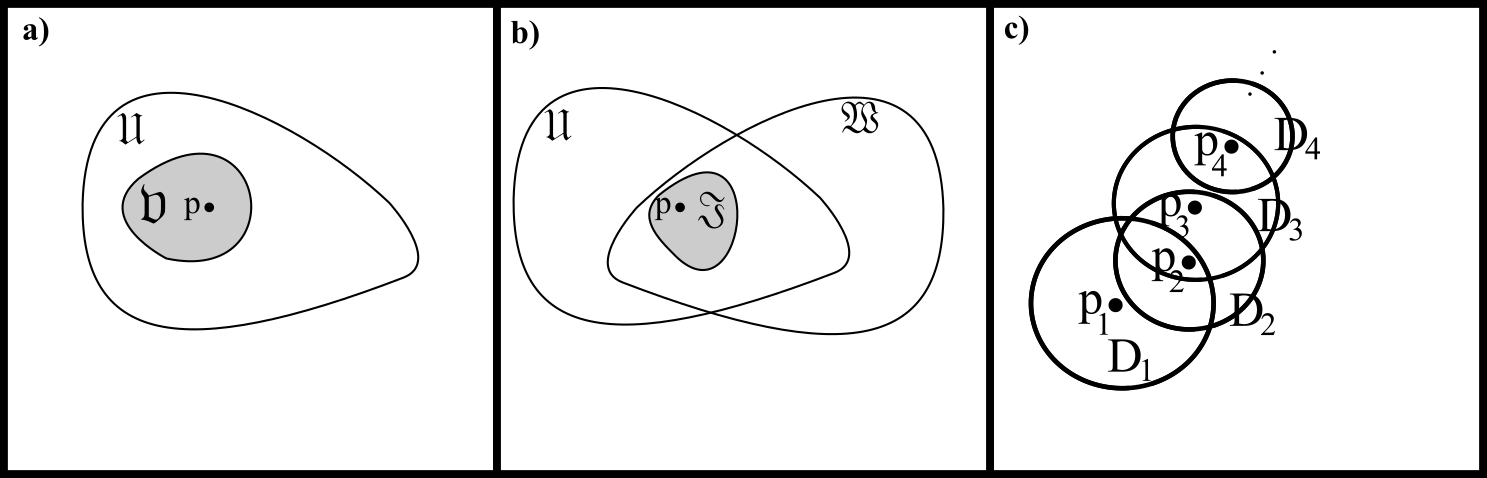}
\caption[Text der im Bilderverzeichnis auftaucht]{\footnotesize{a) Localization: restrict from $\FrU$ to $\FrV \, \subset \, \FrU$.  

\m

\n b) Localization to lie within an intersection of domains or `neighbourhoods' $\FrI \, \subset \, \FrU \, \cap \, \FrW$.    

\m 

\n c) Analytic continuation, as exhibited by the analytic functions, by expanding in a power series valid in a disc $\sD_1$ 
around a point $\sp_1$, then expanding in another power series about an off-centre point $\sp_2 \in \sD_1$ to form a new disc $\sD_2$ 
partly outside $\sD_1$, and so on.}} 
\label{Lie-Assumptions}\end{figure}} 

\subsection{Advent of Topological Spaces}\label{TS}

\n This subsection works toward establishing various mathematical arenas in which Lie's Mathematics is well-defined.

\m 

\n{\bf Definition 1} {\it Topological spaces} \cite{Sutherland, Armstrong, Lee1}\footnote{These are originally due to Hausdorff \cite{Hausdorff}, 
though the below is a subsequent reformulation.} 
\be 
\langle \, \FrX , \, \tau \, \rangle
\ee 
consist of a set $\FrX$ alongside a collection $\tau$ of {\it open subsets},  
\be 
\FrU_{\sfO}  \m , 
\ee
with the following properties. 

\m

\n i) $\FrX, \emptyset \in \tau$.

\m 

\n ii) Closure under arbitrary union: 
\be 
\bigcup_{\sfA} \, \FrU_{\sfA} \, \in \tau  \m ,  
\ee 
for $\fA$ indexing an arbitrary subcollection of the $\fO$.   

\m 

\n iii) Closure under finite intersection: 
\be 
\bigcap_{\sfK = 1}^n  \FrU_{\sfK} \, \in \tau   \m . 
\ee 
\n{\bf Remark 1} For the benefit of less mathematically-experienced readers, topological spaces are a further generalization of metric spaces. 
They represent a further level of abstraction, 
conducted moreover from the point of view of seeing how far many key notions of Analysis, such as convergence and continuity, can be pushed.  
See especially \cite{Sutherland} for a brief and lucid introduction to topological spaces (via metric spaces).
The open subsets are a particular type of collection that is convenient for performing Analysis; so are their complements: the {\it closed sets}.  

\m 

\n{\bf Remark 2} We furthermore use the notion of {\it neighbourhood} $\FrN_{\sx}$ of a point $\mx \in \FrX$, 
meaning a subset of $\FrX$ that contains an open subset $\FrU_{\sx}$ for each of its points $\mx$:  
\be 
\mx \, \in \,  \FrU_{\sx} \m \subseteq \m \FrN_{\sx}  \m .  
\ee 
This is a more specific, modern notion of neighbourhood than Lie's own (hence our use of `neighbourhood' in the preceding subsection).  
It is moreover contemporary practice to not only label domains and neighbourhoods, 
but also to be able to quantify how small these are in the metric space context, 
or, in the general topological space context, their overlap properties and other subset inter-relations.

\m

\n{\bf Remark 3} Topology itself \cite{Armstrong, Munkres} can be quite widely seen as the study of continuity.
The basic notion of continuity in $\mathbb{R}^n$ can be rephrased as preservation of openness under taking inverse images, 
in which form it generalizes to a map between topologial spaces 
\be 
\phi: \langle \, \FrX, \, \tau \, \rangle  \m \longrightarrow \m  \langle \, \FrY , \, \upsilon \, \rangle
\ee 
being {\it continuous} iff the inverse image of each open set is itself open, 
\be 
\FrU \in \upsilon \Leftrightarrows \phi^{-1}(\FrU) \, \in \, \tau \m .
\ee
Topological spaces are thereby indeed a natural setting within which to study topology.  

\m 

\n{\bf Definition 2} A continuous map $\phi$ is furthermore a {\it homeomorphism} if it is a bijection and has continuous inverse.  

\m 

\n{\bf Definition 3} {\it Topological properties} are those properties which are preserved by homeomorphisms.   

\m 

\n{\bf Remark 4} The rest of this subsection outlines the various such that are used in the current Series. 

\m 

\n{\bf Definition 4} Suppose $\FrU_1$ and $ \FrU_2$ are open sets such that 

\m

\n i) $\FrU_1 \, \bigcap \, \FrU_2 = \emptyset$, 

\m 

\n ii) $\FrU_1 \, \bigcup \, \FrU_2 = \FrX$, and 

\m 

\n iii) neither $\FrU_1$ nor $\FrU_2$ are $\emptyset$.  

\m

\n Then $\FrU_1, \FrU_2$ {\it disconnect} $\FrX$.
If $\FrX$ is not disconnected by any two sets, $\FrX$ is {\it connected} \cite{Armstrong, Sutherland, Lee1}. 

\m 

\n{\bf Remark 5} Connectedness is in good part motivated by considering how far the Intermediate Value Theorem \cite{Korner, Sutherland} can be generalized.  

\m 

\n{\bf Definition 5} A {\it path} from point $x$ to point $y$ ($x, y \in \FrX$) is a continuous function 
\be
\iota: \FrI \longrightarrow X 
\ee 
for $\FrI$ the closed unit interval and $\iota(0) = x$, $\iota(1) = y$.  

\m 

\n{\bf Definition 6} A topological space $\FrX$ is {\it path-connected} if any two points $x$, $y$ $\in \FrX$ can be joined by a path. 

\m 

\n{\bf Remark 6} $\mbox{(Path-connectedness)} \m \Rightarrow \m \mbox{(connectedness)}$ but the converse is false \cite{Munkres}.

\m 

\n{\bf Remark 7} Some notions of countability are concurrently topological properties, due to involving counting topologically defined entities.

\m 

\n{\bf Definition 7} {\it First countability} \cite{Lee1, Munkres} holds if for each $x \in \FrX$, 
there is a countable collection of open sets such that every open neighbourhood $\FrN_x$ of $x$ contains at least one member of this collection. 
{\it Second countability} \cite{Lee1, Munkres} is the stronger condition that 
there is a countable collection of open sets such that every open set can be expressed as union of sets in this collection. 
(Such a collection is termed a {\it base}.)   

\m 

\n{\bf Definition 8} Notions of {\it separation} are topological properties which indeed involve separating two objects 
(for instance points, orcertain kinds of subsets) by encasing each in a disjoint subset.  

\m 

\n A particular such is {\it Hausdorffness} \cite{Hausdorff, Sutherland, Lee1}
\n$$
\mbox{for } \mbox{ } \mbox{$x, y \in \FrX \mma  x \neq y \mma  \exists$ \mbox{ } open sets \mbox{ } ${\FrU}_x, {\FrU}_y \in \tau$}
$$
\beq
\mbox{such that \mbox{ } $x \in {\FrU}_x \mma y \in {\FrU}_y$ \m and \m ${\FrU}_x \, \bigcap \, {\FrU}_y  \es  \emptyset$} \m . 
\label{Hausdorffness}
\eeq
{\bf Remark 8} This case thus involves separating points by open sets.  
Hausdorffness allows for each point to have a neighbourhood without stopping any other point from having one.
This is a property of $\mathbb{R}$ and is additionally permissive of much Analysis.
In particular, Hausdorffness secures uniqueness for limits of sequences. 

\m

\n{\bf Definition 9} A collection of open sets   
\be 
\{\FrU_{\sfC}\}
\ee 
is termed an {\it open cover} for $\FrX$ if 
\be 
\FrX  \es  \bigcup_{\sfC} \, \FrU_{\sfC}  \m .
\ee  
On the one hand, a subcollection of an open cover that is still an open cover is termed a {\it subcover}, 
$\{\FrV_{\sfD}\}$ for $\fD$ a subset of the indexing set $\fC$. 

\m 

\n On the other hand, an open cover $\{\FrV_{\sfD}\}$ is a called {\it refinement} of $\{\FrU_{\sfC}\}$ 
if to each $\FrV_{\sfD}$ there corresponds a $\FrU_{\sfC}$ such that 
\be 
\FrV_{\sfD} \m \subset \m \FrU_{\sfC}  \m . 
\ee  
$\{\FrV_{\sfD}\}$ is finally {\it locally finite} if each $x \in \FrX$ has an open neighbourhood $\FrN_x$ such that 
\be 
\mbox{only finitely many } \m \FrV_{\sfD} \m \mbox{ obey } \m  \FrN_x \, \bigcup \, \FrV_{\sfD} \neq \emptyset  \m .  
\ee 
\n{\bf Definition 10} A topological space $\tau(\FrX)$ is {\it compact} \cite{AU29, Armstrong, Sutherland, Lee1} 
if every open cover of $\FrX$ has a finite subcover. 

\m 

\n{\bf Remark 9} Compactness generalizes the important property that continuous functions are bounded on a closed interval of $\mathbb{R}$.

\m 

\n{\bf Definition 11} A topological space $\langle \,  \FrX, \tau\, \rangle$ is {\it locally compact (LC)} \cite{Lee1} 
if each point $\mp \in \FrX$ has a neighbourhood contained in a compact subset $\FrK_{\sp} \m \subseteq \m \FrX$.  
 
\m 

\n{\bf Definition 12} A topological space $\tau(\FrX)$ is {\it paracompact (P)} \cite{Dieu, Lee1} 
if every open cover of $\FrX$ has a locally finite refinement. 

\m

\n{\bf Remark 10} There is fascinating interplay between topological properties: many combinations of these imply other a priori unrelated properties 
(see \cite{Sutherland, Munkres, Lee1} for some basic such, or \cite{Willard, Engelking, Nagata} for more extensive and advanced repertoires).

\m 

\n Except where explicitly stated, we henceforth assume Hausdorffness and second-countability (HS). 
Second countability ensures sequences suffice to probe most topological properties, 
whereas Hausdorffness ensures that neighbourhoods retain many of the intuitive properties of their metric space counterparts.
Along such lines, HS manifests a useful balance between a space being too large, or too small, for the purpose of performing Analysis.  

\m 

\n Hausdorffness extends moreover to how compact sets can be separated by open neighbourhoods, so in Hausdorff spaces `compact sets behave like points'.
$\mathbb{R}^n$ itself not being compact, however, illustrates how for some purposes compactness is too strong a requirement. 
In many cases, LC can effectively deputize; $\mathbb{R}^n$ is moreover LC.  

\m 

\n The first major combination of topological properties that we consider, however, involves the following alongside HS. 

\m 

\n{\bf Definition 13} A topological space $\langle \, \FrX, \tau \, \rangle$ is {\it locally Euclidean (LE)} if every point $x \in \FrX$ has a neighbourhood $\FrN_x$ 
that is homeomorphic to $\mathbb{R}^p$: Euclidean space.

\m 

\n{\bf Definition 14} $\langle \, \FrX, \tau \, \rangle$ is a {\it (real) topological manifold} \cite{Lee1} if it is Hausdorff, second-countable and locally-Euclidean. 

\m 

\n{\bf Remark 11} Let us denote the general topological manifold by 
\be 
\Frm   \mma 
\ee 
and term the above `LEHS' trio of topological space properties `manifoldness'.  
Moreover \cite{Lee1}, 
\be 
\mbox{LEHS} \m \Rightarrow \m \mbox{LC} \m ,  
\ee  
\be 
\mbox{LEHS} \m \Rightarrow \m \mbox{P}  \m ,   
\ee    
and that, for LEHS, $\mbox{(path-connectedness)} \m \Leftrightarrow \m \mbox{(connectedess)}$.
The last of thes justifies our more modern interpretation being termed `(path-)connected' in the preceding subsection.  

\m 

\n{\bf Definition 15} A {\it topological group} \cite{AMP, Lee1} is a set equipped with both a topology 
and a group operation such that the composition and inverse operations are continuous.
Sec \ref{Lie-Groups}'s Lie groups -- which are groups that are also topological manifolds -- are a major example of this.

\subsection{Context of contemporary Theoretical Physics}\label{SAU}
%
{\begin{figure}[!ht]
\centering
\includegraphics[width=0.66\textwidth]{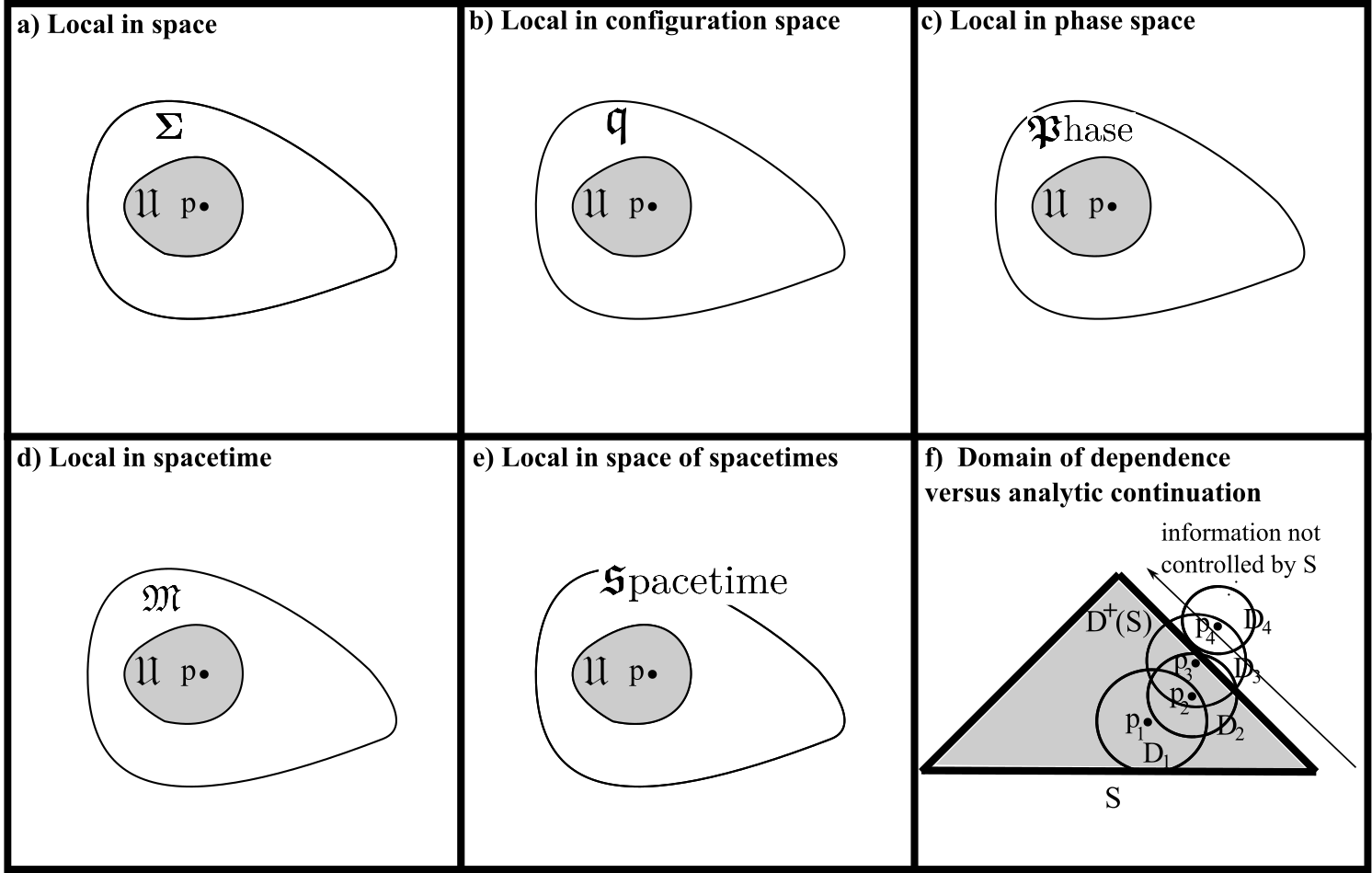}
\caption[Text der im Bilderverzeichnis auftaucht]{\footnotesize{Local in: a) space, b) configuration space, c) phase space, d) spacetime and e) space of spacetimes.

\m

\n f) Analytic continuation does not respect the domain of dependence property.}} 
\label{Mod-Phys-Settings}\end{figure}} 

\n{\bf Structure 1} Contemporary Theoretical Physics also necessitates extending our mathematical setting to state spaces $\bFrS(\lFrs)$ for a system $\lFrs$ such as 
configuration space $\FrQ(\lFrs)$, 
configuration-velocity space $\FrQ\bFrV(\lFrs)$, 
                                                                   phase space $\Phase(\lFrs)$, 
                                                                                                             and the space of spacetimes $\Spacetime(\lFrs)$. 
These state spaces have, respectively, 
$\bfQ$, 
$(\bfQ, \dot{\bfQ})$, 
$(\bfQ, \bfP)$, 
$\bfS$ as {\it base objects}, 
which we denote collectively by $\bfB$; 
we denote the general case of state space by forming a state space $\bFrS$.  
Locality can also be in each of these state spaces, whereas multiple applications of locality within a given working can apply in multiple settings concurrently. 

\m 

\n{\bf Structure 2} Some applications furthermore require generalized objects as function(al)s of base objects, $\bfO(\bfB)$, forming spaces of objects $\bFrO$.  
 
\m
																																																					  
\n{\bf Structure 3} Let us also introduce the notion of {\it encoder functions} $\cE$ as a suitable generalization of the familiar Hamiltonians $\cH$ and Lagrangians $\cL$ 
(the latter admitting moreover both spacetime and split space-time forms). 
Space without encoder function is a matter of Geometry, while spacetime without encoder function is a matter of spacetime Geometry rather than spacetime Physics. 
This is how our study picks up both purely geometrical and physical branches.
Note also the distinction between Lagrangians of matter fields on spacetime, Lagrangians of spacetime, and Lagrangians of spacetime with matter fields.   

\m

\n{\bf Remark 1} Analyticity is out due to not respecting Relativistic Physics' causal notions; 
more concretely, analyticity does not know to respect the Domain of Dependence property \cite{Wald} (Fig \ref{Mod-Phys-Settings}.b).

\m  
 
\n{\bf Remark 2} Even though canonical formulations only began to see widespread use \cite{Dirac26, Dirac30} with the advent of QM in the 1920s, 
Lie nonetheless already did consider \cite{Landsman} what can be recognized to be momentum maps and Poisson algebraic structures around 1890. 
Dirac's other great use of canonical formulations \cite{Dirac51, Dirac58, Dirac} -- to study constrained systems -- was 60 to 70 years after Lie's time.   
Such details of Lie's work however fell into obscurity, by which momentum maps and Poisson algebraic structures 
themselves had to await rediscovery in the 1960s.  
Substantial work was built on these in the 1970s \cite{MW74} by which they have become mainstream \cite{Vaisman, Gengoux}.    
Substantial other amends to Lie Theory of relevance in this Series also largely only started to appear at this point \cite{Serre-Lie, G64, NR66}. 

\m 

\n{\bf Remark 3} All in all, some aspects of Lie's work were, mathematically, around 80 or 90 years ahead of his time.   
Pace \'{E}lie Cartan \cite{Cartan55}, who significantly extended Lie's work in the 1900s through to the 1930s, 
it took around 80 or 90 years for much of Lie's work to be rediscovered or elsewise substantially extended upon.    
And pace Dirac, who had a 2-decade start on everybody else within the more restricted setting of constraint algebraic structures 
and the associated observables \cite{DiracObs, Dirac51}.

\subsection{Context of Differential Geometry}

\n Flat geometries $\mathbb{R}^d$ (or $\mathbb{C}^m$) constitute the traditional arena for Geometry, 
to which the current Article contributes some Foundations of Geometry material.
Working with manifolds $\Frm^d$ is more general.
At least the simpler state spaces can also be taken to be manifolds (in unreduced or fortunately-reduced cases).

\m 
 
\n{\bf Remark 1} Sec \ref{Lie-Modelling-1}'s structures -- PDEs, flows, Lie dragging, Killing's Mathematics -- carry over to this setting \cite{Yano55, Yano70, Lee2}. 

\m 
 
\n{\bf Remark 2} Real manifolds are most usually considered to be {\it smooth} -- $\FrC^{\infty}$ -- 
or sufficiently differentiable, which for physical applications, usually means $\FrC^2$: {\it twice continuously differentiable functions}.   
In fact, aside from the analytic functions and the merely {\it continuous functions} $\FrC^0$ behaving qualitatively differently 
in both Differential-Geometric and Relativistic Physics terms by which they are to be excluded, 
neither Differential Geometry \cite{Whitney36} nor Relativistic Physics \cite{Wald} are particularly sensitive to further details of which function space is in use.  
Within $\FrC^k$ function spaces -- {\it k times continuously differentiable functions} -- any $k \geq 2$ will do (including $k = \infty$). 
More modern and advanced work moreover often employs instead \cite{HE73, R08, CB} (perhaps weighted) Sobolev spaces \cite{Dafermos} on manifolds. 

\m 

\n We consider the Lie group operation to be correspondingly smooth, 
giving a convenient and more contemporary standard (Lie having restricted detailed attention to the analytic case).

\subsection{Lie's assumptions updated}\label{Update}

Overall, we drop two of Lie's modelling assumptions but keep and expand upon the remaining one, as follows. 

\m 

\n{\bf Update 1} We eschew analyticity, by which Lie's style of proof will not do. 
This amounts to needing to update to a somewhat harder class of function spaces, 
within which setting further changing exactly which function space is involved makes little conceptual difference. 

\m 

\n{\bf Update 2} We keep the idea of free reallocation: the `local' in ALRoPoT is of this kind, 
in fact extended to hold over a series of auxiliary spaces as well as to the underlying space or spacetime setting in question.

\m 

\n{\bf Update 3} We do specify particular domains and neighbourhoods, because of the development of topological spaces since Lie's era and to ask more precise questions.

\section{Differentiable Manifolds}\label{Diff-Manifold}

\subsection{Meshing conditions and atlases}\label{Mesh}

\n{\bf Structure 1} Study of manifolds $\Frm$ benefits from Riemann's notion of {\it chart} alias {\it local coordinate system} for $\Frm$ (Fig \ref{Top-Man}.a): an injective map 
\be 
\varphi: \FrU \longrightarrow \varphi(\FrU) \m \subset \m \mathbb{R}^n
\ee 
for $\FrU$ an open subset of $\Frm$.  
Each chart does not in general cover the whole manifold; one gets around this by considering a suitable collection of charts.
These serve as homeomorphisms which guarantee the locally Euclidean property.  
%

\m 

\n{\bf Structure 2} One is to compare those charts which overlap, leading to the two-chart Fig \ref{Top-Man}.b), with 
\be
\varphi_1: \FrU_1 \longrightarrow \mathbb{R}^n  \mma 
\ee  
\be                                                                                                
\varphi_2: \FrU_2 \longrightarrow \mathbb{R}^n 
\ee 
which do indeed overlap: 
\be 
\FrU_1 \, \bigcup \, \FrU_2 \m \neq \m 0  \m .
\ee  
\n{\bf Structure 3} Let us next consider a composite map 
\beq 
t_{12} \:= \varphi_2 \circ \varphi_1^{-1}
\label{transition}
\eeq 
sends $\FrU_1 \, \bigcup \, \FrU_2$ to itself.
This is a locally defined map $\mathbb{R}^n \longrightarrow \mathbb{R}^n$; it is a local coordinate transformation, and is called a {\it transition function}.
An {\it atlas} for a topological manifold is a collection of charts that, between them, cover the whole manifold. 
%
{\begin{figure}[!ht]
\centering
\includegraphics[width=0.5\textwidth]{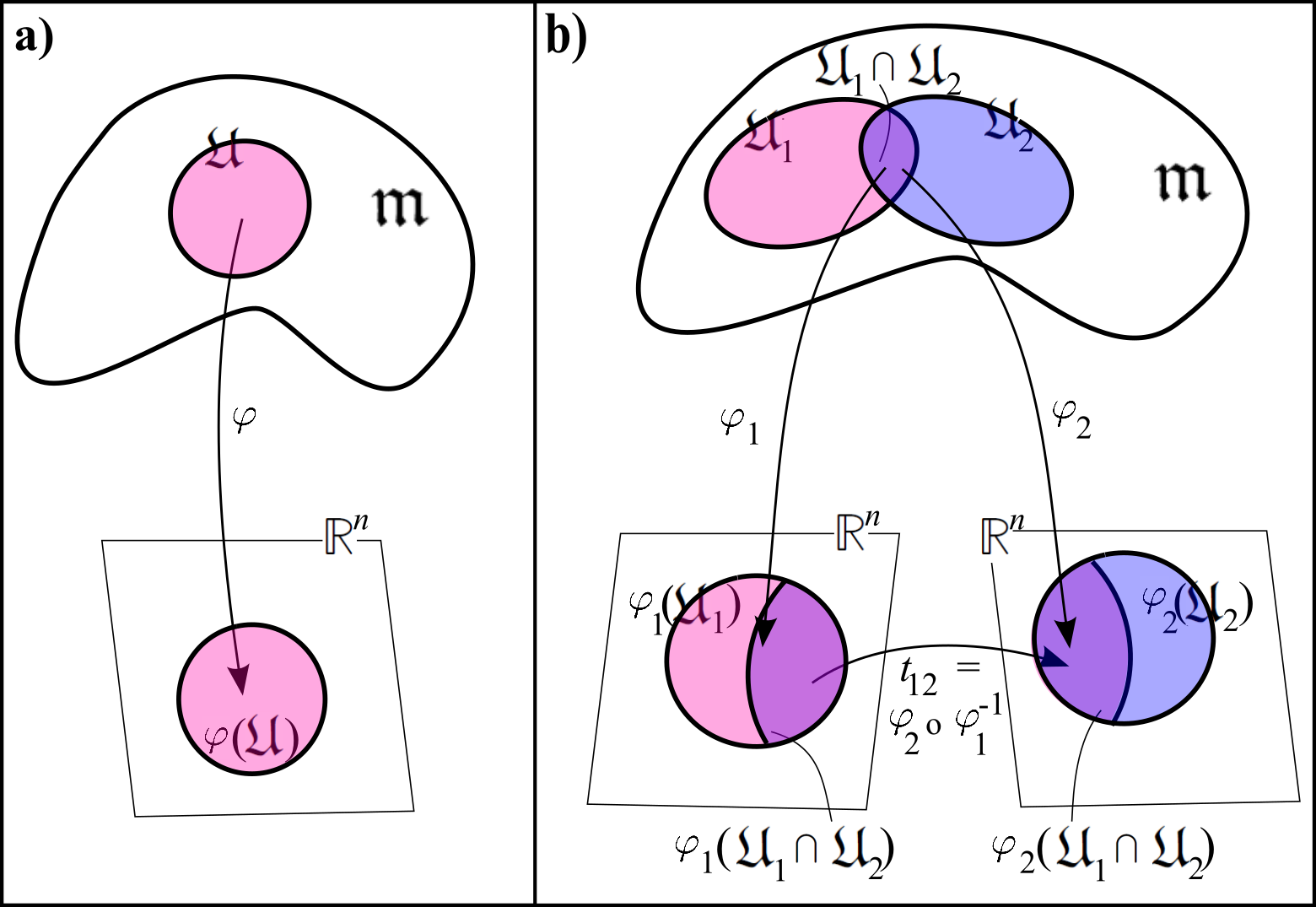}
\caption[Text der im Bilderverzeichnis auftaucht]{\footnotesize{a) A chart.  

\m 

\n b) Overlapping charts and transition functions $t_{12}$.}} 
\label{Top-Man}\end{figure}} 

\m 
 
\n{\bf Structure 4} Charts can furthermore allow for one to tap into the standard $\mathbb{R}^p \longrightarrow \mathbb{R}^q$ Calculus 
(as supported by the Analysis that is rooted on manifolds being HS).  
This allows for manifolds to be equipped with {\it differentiable structure} \cite{Lee2, KN1} in addition to topological structure. 

\m 

\n{\bf Structure 5} Such {\it differentiable manifolds} possess not only a local differentiable structure in each coordinate patch $\FrU_i$ 
but also a notion of global differentiable structure.  
This is due to the `{\it meshing condition}' on the coordinate patch overlaps (Fig \ref{Top-Man}.b).
In this setting, the transition functions be interpreted as 
Jacobian matrices of derivatives for one local coordinate system $\bx$ with respect to another $\bar{\bx}$:   
\beq
{\mL^A}_B  \es  \frac{\pa (x^A)}{\pa(\bar{x}^B)} \m .
\eeq  
[We use capital Latin indices on the general manifold $\Frm$.]

\m 

\n{\bf Structure 6} The above topological manifold notion of atlas can also be equipped with differentiable structure. 
Our main interest here is moreover really in equivalence classes of atlases; 
differentiable structure is then often in practice approached using a convenient small atlas \cite{Nakahara}. 

\m 

\n{\bf Remark 1} Having Calculus available throughout the manifold, moreover, allows on to study {\it differential equations} 
which can in turn represent Physical Law in a conventional manner.

\subsection{Vectors and tensors}\label{VT}
%
{\begin{figure}[!ht]
\centering
\includegraphics[width=0.85\textwidth]{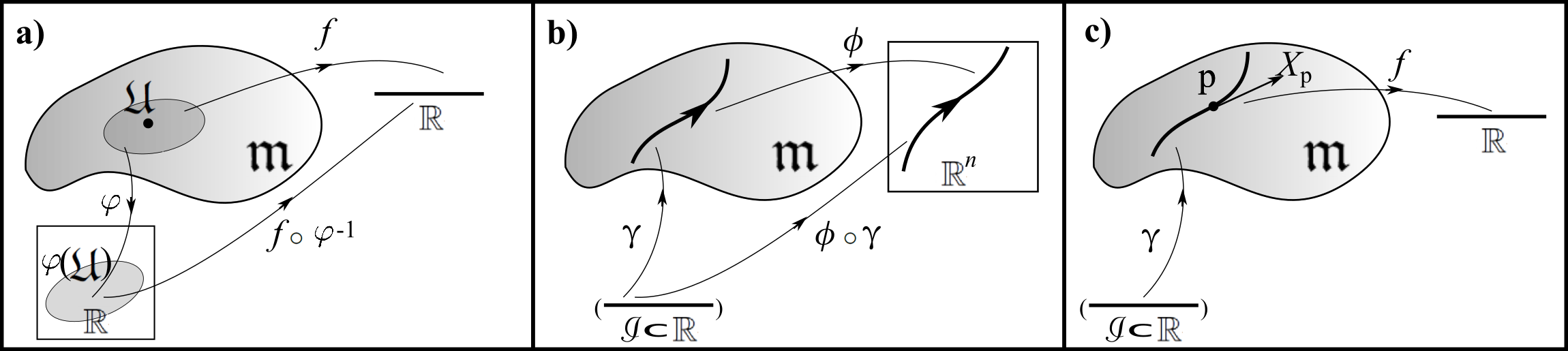}
\caption[Text der im Bilderverzeichnis auftaucht]{\footnotesize{a) A function on a manifold. 

\m 

\n b) The curve construct on a manifold.  

\m 

\n c) A notion of vector on a manifold based on the curve construct and action on a fiducial function.}} 
\label{Diff-Man}\end{figure}} 

\n{\bf Structure 1} {\it Functions} on manifolds are defined as per Fig \ref{Diff-Man}.a).  

\m 

\n{\bf Structure 2} Let us next introduce {\it vectors} on manifolds as the tangents to {\it curves}, which are themselves mappings 
\be 
\FrI \longrightarrow \Frm
\ee  
for $\FrI  \m  \subset \m \mathbb{R}$ a closed interval (as per Fig \ref{Diff-Man}.b); compare also Sec \ref{TS}'s notion of path). 
The vectors themselves are maps (\cite{Stewart} and Fig \ref{Diff-Man}.c)   
$$ 
\gamma^{\prime}_{\sp} : \m \FrC^{\infty}(\mathbb{R}) \m \longrightarrow \m \mathbb{R}
$$
\be 
f: \m \m  \mapsto \m \m \left. \frac{\d }{\d \nu} f \circ \gamma \right|_{\nu = 0}  \m .  
\ee 
\n{\bf Structure 3} The vectors thus defined at a given point $\mp$ form the tangent space at $\mp$, 
\be 
\FrT_{\sp}(\Frm)   \m .  
\ee 
\n{\bf Remark 1} One can furthermore compose curve and chart maps to make use of standard $\mathbb{R}^p \longrightarrow \mathbb{R}^q$ Calculus.     

\m 

\n{\bf Remark 2} One can additionally straightforwardly show that all notions involved are chart-independent: a well-definedness criterion \cite{Stewart}.  

\m 

\n{\bf Remark 3} One can finally apply \cite{KN1, Wald, Stewart} the basic machinery of Linear Algebra to produce the following notions.  

\m 

\n{\bf Structure 4} At a point $\mp$ on the manifold, a covector is a linear map 
\be 
\FrT_{\sp}(\Frm) \longrightarrow \mathbb{R}  \m . 
\ee
\n{\bf Structure 5} The covectors at $\mp$ form the cotangent space 
\be 
\FrT_{\sp}^*(\Frm) \m : 
\ee 
the Linear Algebra dual of the tangent space.   

\m 

\n{\bf Structure 6} The {\it rank $(k, \, l)$ tensors} at $\mp$ are multilinear maps 
\be 			   
\bigtimes_{i = 1}^k \FrT_{\sp}^*(\mbox{{\sl \Frm}}) \times \bigtimes_{j = 1}^l \FrT_{\sp}(\mbox{{\sl \Frm}}) \m \longrightarrow \m  \mathbb{R}  \m .
\ee
\n{\bf Structure 7} A union of vectors, one at each $\mp \, \in \, \Frm$, constitutes a {\it vector field}  over $\Frm$; 
                                                                                        {\it tensor fields} are similarly defined.  
In terms of components, $(k, \, l)$-tensors transform according to 
\be
{\mT^{\bar{A}_1 \, ... \, \bar{A}_k}}_{\bar{B}_1 \, ... \, \bar{B}_l}  \es   
{\mL^{\bar{A}_1}}_{A_1} \, ... \, {\mL^{\bar{A}_k}}_{A_k}{\mL^{B_1}}_{\bar{B}_1} \, ... \, 
{\mL^{B_l}}_{\bar{B}_l} {\mT^{A_1 \, ... \, A_k}}_{B_1 \, ... \, B_l}
\label{tentranslaw}
\ee
in passing between plain and barred coordinate systems.  

\m 

\n{\bf Structure 8} Let us use $\bG$ more geometrical objects with given transformation laws for a wider range than just tensors, 
including e.g. also densities \cite{Wald} and connections \cite{KN1}.  

\section{Lie derivatives}\label{Lie-Deriv}

\subsection{Notions of derivative}  

\n Physics and Differential Geometry make plentiful use of {\it derivatives} acting on $\bG$.  

\m 

\n On the one hand, such are not straightforward to set up in generally curved geometry.
For the flat-space derivatives that one is accustomed to entail taking the limit of the difference between vectors at different points.  
However, in the context of differentiable manifolds, such vectors belong to different tangent spaces.  
Whereas in $\mathbb{R}$ one can just move the vectors to the same point, there is no direct counterpart of this procedure on a general manifold 
(c.f.\ Fig \ref{Diff-Man}.c).  
The usual partial derivation is undesirable since it does not preserve {\it tensoriality}: the mapping of tensors to tensors.  

\m 

\n On the other hand, for action on tensors, it suffices to construct such a notion of derivatives acting on vectors and acting trivially on scalars. 
This is because the derivative's action on all the other tensors can then be found by application of the Leibniz rule.

\subsection{Diffeomorphisms}\label{Diffeos}

\n{\bf Definition 1} A {\it diffeomorphism} (see in particular \cite{Lee2}) is a map 
\be 
\phi: \m \Frm \m \longrightarrow \m  \Frm
\label{Diff}
\ee 
that is injective, $\Frc^{\infty}$ (or e.g.\ $\Frc^k$), and has a an inverse map of matching minimal standard of differentiability.   
These are differentiable manifolds' corresponding automorphisms, forming the group  
\be 
Aut(\Frm)  =  Diff(\Frm)     \m .  
\label{Diff-Group}
\ee
\n{\bf Structure 1} The map (\ref{Diff}) induces a {\it push-forward} (Fig \ref{Map-Push-Pull}.b) on the tangent space
\be 
\phi_* \m : \m \m \FrT_{\sp}(\Frm) \m \longrightarrow \m \FrT_{\phi(\sp)}(\Frm)
\ee  
which maps the tangent vector to a curve $\upgamma$ at $\mp$ to that at the image of the curve $\phi(\upgamma)$ at $\phi(\mp)$.  

\m 

\n{\bf Structure 2} It also induces a {\it pull-back} (Fig \ref{Map-Push-Pull}.c) on the cotangent space 
\be 
\phi^* \m : \m \m \FrT_{\phi(\sp)}^*(\Frm) \m \longrightarrow \m \FrT_{\sp}^*(\Frm)
\ee 
which maps 1-forms in the opposite direction.  
%
{\begin{figure}[!ht]
\centering
\includegraphics[width=0.85\textwidth]{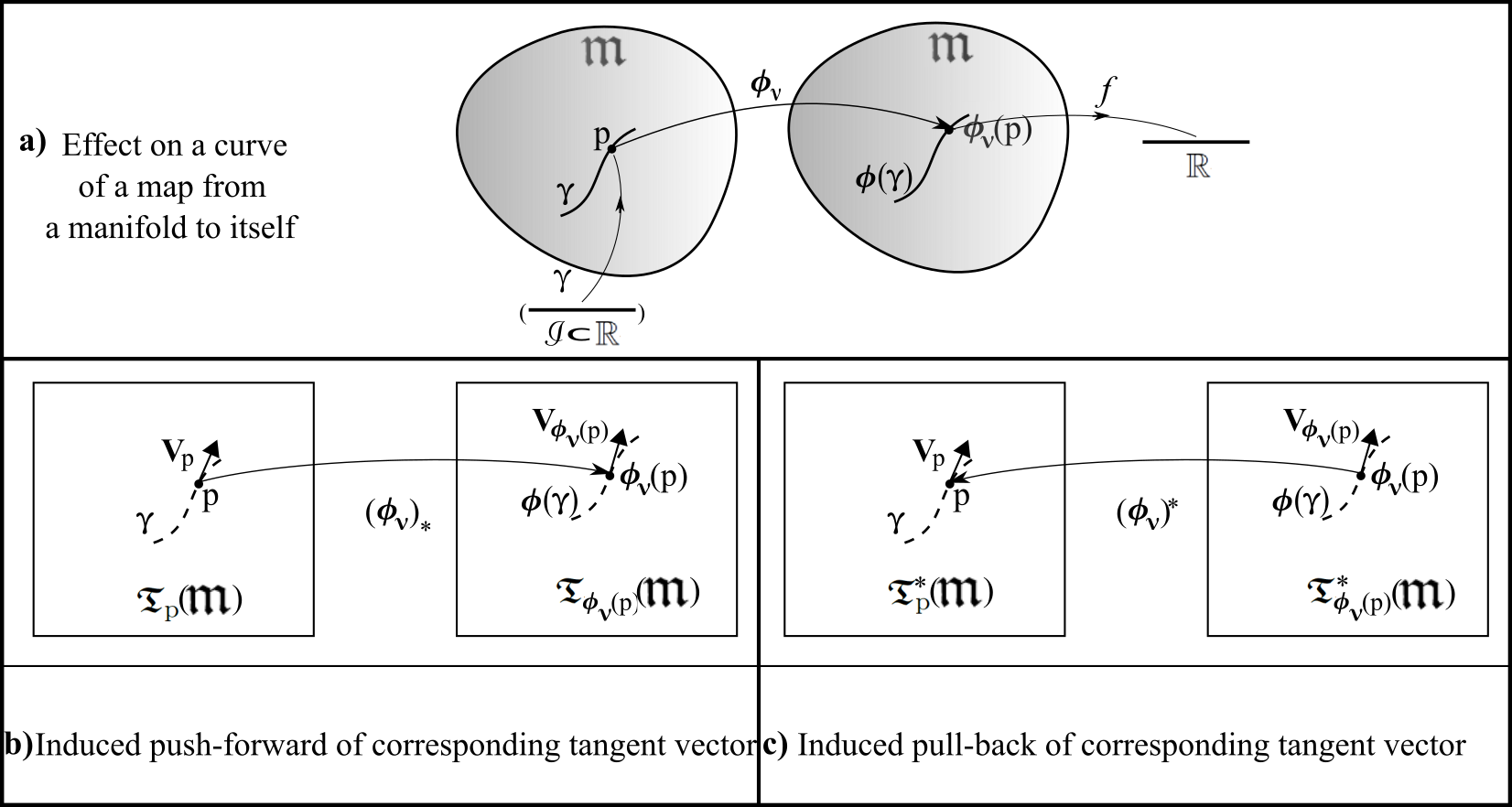}
\caption[Text der im Bilderverzeichnis auftaucht]{\footnotesize{a) Effect on a curve of mapping a manifold to itself. 
b) The induced push-forward and c) pull-back (used in Article XII's consideration of foliations).} } 
\label{Map-Push-Pull}\end{figure}} 

\m 

\n{\bf Structure 3} 
\be 
\phi_* \bG  =  \bG
\ee 
defines a {\it symmetry} for the general geometrical object $\bG$.

\subsection{Integral curves}\label{IC}

\n{\bf Definition 1} An {\it integral curve} (see e.g.\ \cite{Stewart}) of a vector field $\bV$ in a manifold $\Frm$ 
is a curve $\upgamma(\nu)$ such that the tangent vector is $\bV_{\sp}$ at each $\mp$ on $\upgamma$ (Fig \ref{IC-Com}.a). 

\m 

\n{\bf Remark 1} These have local existence-and-uniqueness by standard ODE theory \cite{Lee2}.  

\m 

\n{\bf Definition 2} A set of complete integral curves corresponding to a non-vanishing vector field is called a {\it congruence}. 

\m 

\n{\bf Remark 2} This `fills' a manifold or region therein upon which the vector field is non-vanishing: the curves go through all points therein.
A second interpretation of flow is as a congruence of integral curves.   
The 1-parameter subgroup's generator for a flow $\gamma(\nu)$ is moreover the tangent vector $\gamma^{\prime}(0)$. 

\m 

\n{\bf Structure 1} For later reference, proceeding along two local congruences of integral curves in either order (Fig \ref{IC-Com}.b) produces, to leading order, 
the commutator 
\be 
x^i_{\sv} - x^i_{\su} \m = \m  [ \, \biX, \, \biY \, ] \, \d\mu \, \d\nu  \m + \m  O(\d^3)  \m . 
\ee 
%
{\begin{figure}[!ht]
\centering
\includegraphics[width=0.35\textwidth]{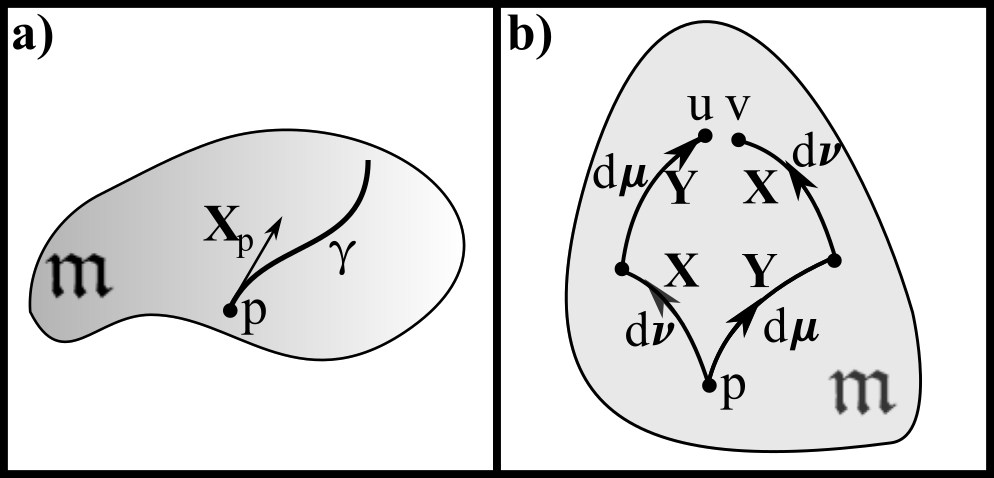}
\caption[Text der im Bilderverzeichnis auftaucht]{\footnotesize{a) Integral curve on a manifold. 
b) Commutator corresponding to proceeding along two local congruences of integral curves in either order.} } 
\label{IC-Com}\end{figure}} 

\subsection{Lie derivatives}\label{Lie-Deriv-Sub}
%
{\begin{figure}[!ht]
\centering
\includegraphics[width=0.85\textwidth]{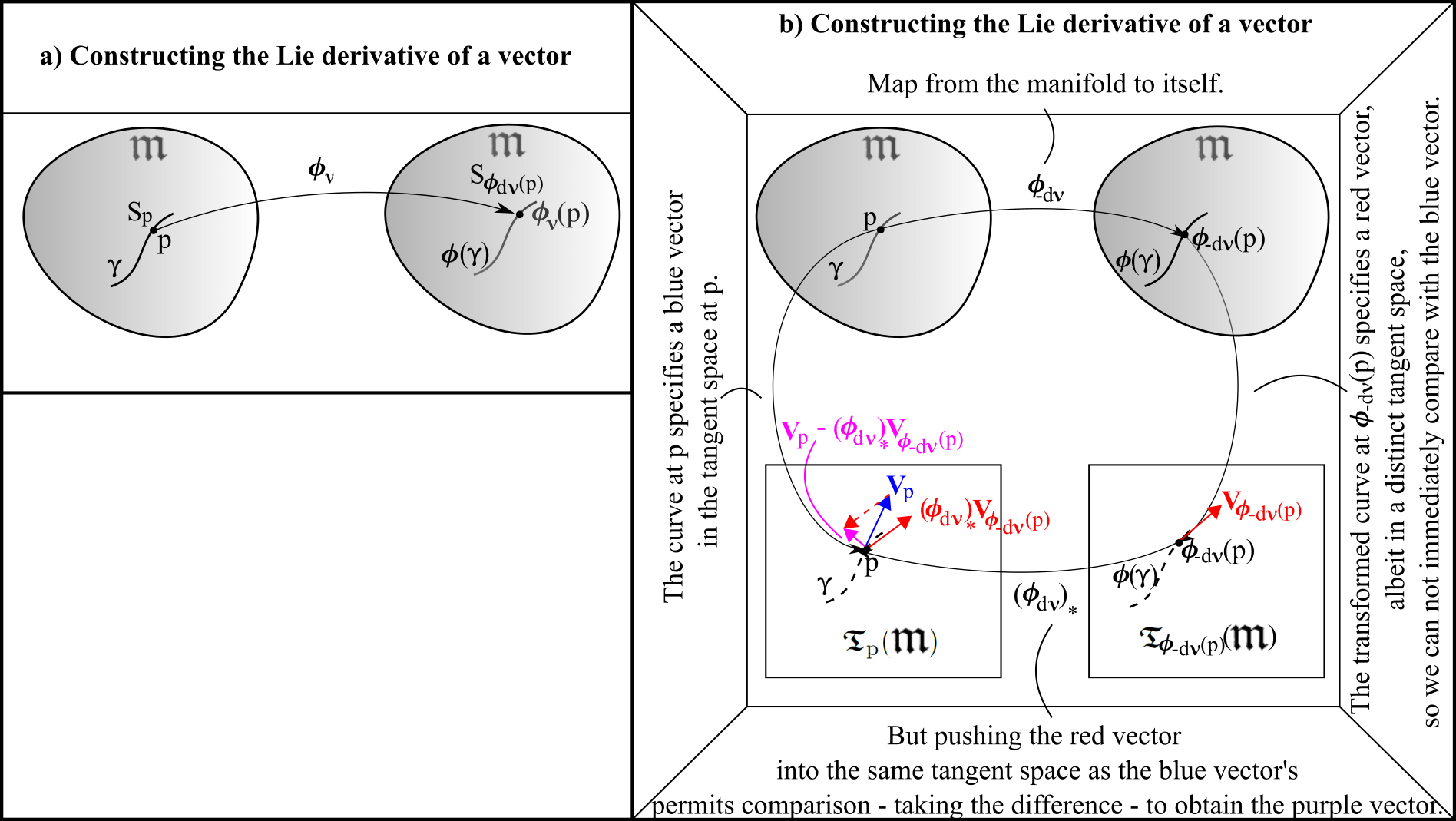}
\caption[Text der im Bilderverzeichnis auftaucht]{\footnotesize{Decomposition of the first-principles construction of the Lie derivative of 
a) a scalar and 
b) a vector.} } 
\label{Lie-19}\end{figure}} 

\n{\bf Structure 1} First-principles considerations using Fig \ref{Lie-19}.a)-b)'s constructs 
give the actions of Lie derivation on scalars and vectors as the first equalities below. 
For $\upgamma$ the integral curve of $\bupxi$ through $\mp$ inducing a 1-parameter group of transformations $\phi_{\nu}$ with parameter $\nu$, 
the Lie derivative with respect to $\bupxi$      at $\mp$ of a scalar $\mS$ is 
\be
(\pounds_{\sbupxi} \mS )_{\sp}  \es  \m \stackrel{\mbox{lim}}{\d \nu \rightarrow 0} 
\left(
\frac{\mS_{\phi_{\d \nu} (\sp)} - \mS_{\sp}}{\d \nu}                                                           \m .
\right)
\ee
For a vector $\bV$, it is 
\be
(\pounds_{\sbupxi}\mV^A)_{\sp} = \m \stackrel{\mbox{lim}}{\d \nu \rightarrow 0} 
\left(
\frac{\mV^A_{\sp} - (\phi_{\d \nu})_* \mV^A_{\phi_{-\sd \nu}(\sp))}}{\d \nu} 
\right)                                                                                                        \m .
\ee
Consult e.g.\ \cite{Stewart} as regards using the `useful Lemma' to pass to the following`computational' forms in each case:    
\be 
\pounds_{\sbupxi} \mS    \es  \xi^A \pa_A \mS                                                                 \m , 
\label{slie}
\ee
and 
\be  
\pounds_{\sbupxi} \mV^A  \es  \xi^B \pa_B \mV^A - \pa_B {\xi^A} \mV^B 
                                     \es  [ \, \bupxi , \, \bV \, ]^A                                         \m .  
\ee
The latter gives the differential-geometric commutator, 
which can in turn be interpreted in terms of advancing along two different pairs of integral curves \cite{Stewart} as per Fig \ref{IC-Com}.b).    

\m

\n One can then readily obtain the Lie derivatives for tensors \cite{Yano55} of all the other ranks from these scalar and vector results by use of Leibniz's rule.

\m 

\n{\bf Remark 1} As a derivative, the Lie derivative is tensorial, 
and {\it directional} in the sense of involving an additional vector field $\bxi$ along which the tensors are dragged.  

\m 

\n{\bf Remark 2} Lie derivatives generate the local infinitesimal version of the diffeomorphisms.   

\m 

\n{\bf Remark 3} {\it Lie dragging} involves moving an object along a particular vector field's (or equivalently flow's) integral curves, 
by means of the Lie derivative with respect to the corresponding vectors.

\section{Lie algebras and Lie groups}\label{Lie-Algebra}

\subsection{Lie algebras}

\n{\bf Definition 1} A {\it Lie algebra} 
\be 
\Frg 
\ee 
is a vector space equipped with a {\it Lie bracket} product: a bilinear map  
\be 
\mbox{\bf |[} \m \mbox{\bf ,} \, \m \, \mbox{\bf ]|} \m : \m \m   \Frg \times \Frg \longrightarrow \Frg 
\label{Lie-Bra}
\ee
that is antisymmetric 
\be 
\mbox{\bf |[} \,   g \mbox{\bf ,} \, h \, \mbox{\bf ]|}   \es   - \mbox{\bf |[} \,  h \mbox{\bf ,} \, g \, \mbox{\bf ]|}  
\m \m  \forall \m g, \m h \, \in \,  \Frg                                                                                         \m .
\label{Anti}
\ee 
and obeys the {\it Jacobi identity}     
\beq
\mbox{\bf |[} \, g \mbox{\bf ,} \, \mbox{\bf |[} \, h \mbox{\bf ,} \, k  \, \, \mbox{\bf ]|} \, \, \mbox{\bf ]|}  \m + \m  \mbox{cycles}  \es  0 
\label{Jacobi-id}
\m \m \forall \m  g, \m h, \m  k \m \in \, \Frg                                                                                   \m .
\eeq
\n{\bf Remark 1} This a subcase of {\it algebraic structure}: equipping a set with a second or further product operations.  

\m 

\n{\bf Example 1} The familiar Poisson brackets 
\be 
\mbox{\bf \{} \m \mbox{\bf ,} \, \m \mbox{\bf \}}
\ee 
are Lie brackets that are additionally a {\it derivation}, i.e.\ obeying the {\it Leibniz} alias {\it product rule}, 
\be 
\mbox{\bf \{} \, A \mbox{\bf ,} \, B \, C \, \mbox{\bf \}}  \es  B \, \mbox{\bf \{} \, A \mbox{\bf ,} \, C \, \mbox{\bf \}}       \m + \m 
                                                                      \mbox{\bf \{} \, A \mbox{\bf ,} \, B \, \mbox{\bf \}} \, C    \m .     
\ee
\n{\bf Example 2} Quantum commutators are also Lie brackets.  

\m 

\n{\bf Definition 2} A spanning set (most efficiently a basis) of elements for a Lie algebra are termed {\it generators}. 
Let us denote these by 
\be 
\uc{\scg}  \mma \mbox{indexex by } \m \fG  \m . 
\label{Gen}
\ee 
\n{\bf Remark 2} Given a basis of generators $\scg$, computing  
\beq
\mbox{\bf |[} \, \scg_{\sfG} \mbox{\bf ,} \, \scg_{\sfG^{\prime}} \, \mbox{\bf ]|}  \es  {G^{\sfG^{\prime\prime}}}_{\sfG\sfG^{\prime}} \scg_{\sfG^{\prime\prime}}                          \m ,
\label{Str-Const}
\eeq 
permits us to read off the {\it structure constants} ${G^{\sfG^{\prime\prime}}}_{\sfG\sfG^{\prime}}$ for the Lie algebra with respect to this basis.  

\m 

\n This can be recast as \cite{III}
\be
\mbox{\bf |[} \,  \uc{\scg} \mbox{\bf ,} \,  \uc{\scg}^{\prime} \, \mbox{\bf ]|}   \es   \uc{\uc{\uc{\biG}}} \, \uc{\scg}^{\prime\prime}      \m , 
\label{L-Algebra}
\ee 
for $\biG$ {\it structure constant 3-arrays} or {\it trilinear maps}: a more succinct and coordinate-independent presentation.

\m 

\n It readily follows from (\ref{Str-Const}, \ref{Anti}) that the structure constants obey antisymmetry 
\be 
{G^{\sfG^{\prime\prime}}}_{\sfG\sfG^{\prime}}  \es  - {G^{\sfG^{\prime\prime}}}_{\sfG^{\prime}\sfG}                                  \m , 
\ee 
and from (\ref{Str-Const}, \ref{Jacobi-id}), the homogeneous-quadratic restriction
\be
{G^{\sfG}}_{ [ \sfG^{\prime}\sfG^{\prime\prime}}{C^{\sfG^{\prime\prime\prime}}}_{\sfG^{\prime\prime\prime\prime} ] \sfG}  \es  0     \m . 
\label{firstJac}
\ee  
%
{            \begin{figure}[!ht]
\centering
\includegraphics[width=0.7\textwidth]{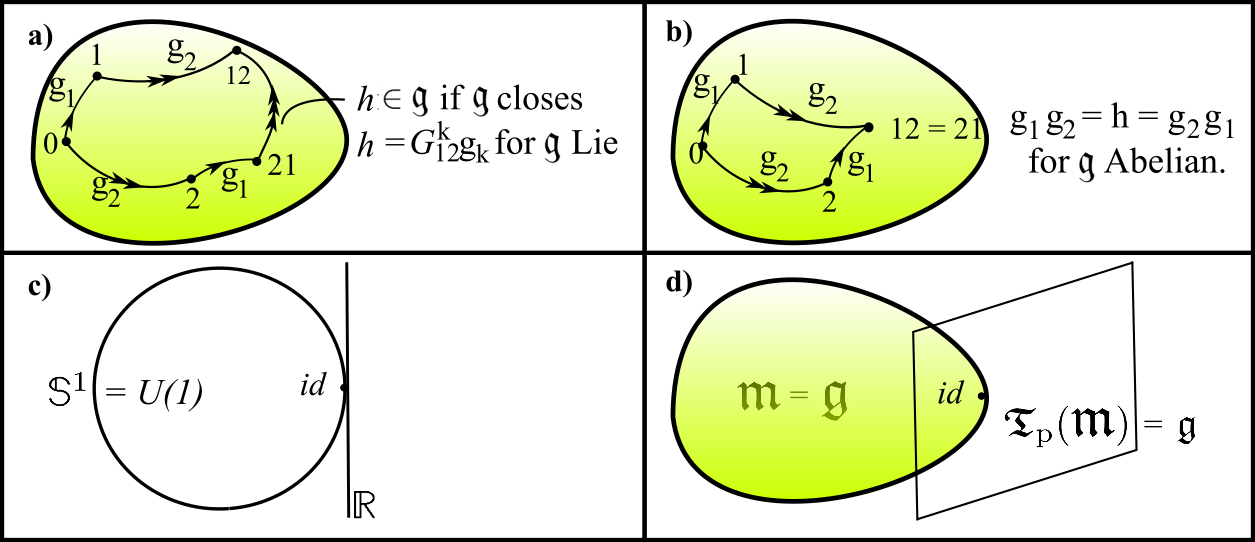}
\caption[Text der im Bilderverzeichnis auftaucht]{        \footnotesize{a) An algebra's commutator. 
This compares applying two transformations $g_1$, $g_2$ in either order to a common initial object $0$.  

\m 

\n b) The even more straightforward commuting subcase, for which the final objects $12$ and $21$ coincide as well.  
Many instances of a) and b) occur in ALRoPoT, 
as picked out among the Series's figures by being depicted on lime-green egg-shaped spaces.

\m 

\n c) The real line as tangent to $U(1)$'s $\mathbb{S}^1$ and 
   d) the general Lie algebra as tangent space to the identity of the general Lie group.} }
\label{Group-Lie-Abelian} \end{figure}          }

\subsection{Lie groups}\label{Lie-Groups}

{\bf Definition 1} {\it Lie groups} $\lFrg$ \cite{Gilmore, Serre-Lie, BCHall} are groups that are concurrently differentiable manifolds; 
additionally their composition and inverse operations are matchingly differentiable. 

\m 

\n{\bf Structure 1} Finite transformations form Lie groups; 
large transformations are included among these in the mixed case; 
see Sec \ref{Basic-Ex} for examples.

\subsection{From Lie groups to Lie algebras}\label{Lie-Lie}

\n{\bf Structure 1} We can view the line $\mathbb{R}$ realized as $i \, \theta$ in $\mathbb{C}$ as (Fig \ref{Group-Lie-Abelian}.c) the tangent to 
\be 
\{z \, \in \, \mathbb{C}  \, | \, |z| = 1\}  \es  \mathbb{S}^1  
                                              =           U(1)   \m . 
\ee 
This corresponds to the exponential map 
\be 
\theta \m \longrightarrow \m \mbox{exp}(i \, \theta)
\ee 
which is valid locally, i.e.\ for a sufficiently small interval of $\mathbb{R}$ (less than $2 \, \pi$ in length).  

\m

\n{\bf Structure 2} A tangent space interpretation continues to apply 
\cite{C46} in the case of higher-dimensional Lie groups $\lFrg$ (Fig \ref{Group-Lie-Abelian}.d), 
by which the corresponding Lie algebras $\Frg$ can be viewed as `tangent space' near $\lFrg$'s identity element. 
This can be set up by considering 1-parameter subgroups \cite{Eisenhart33, Yano55, Stewart, Lee2} (a reinterpretation of integral curves) one at a time.

\m 

\n{\bf Remark 1} The Lie algebra is, on the one hand, more straightforward to handle than a Lie group, 
out of being a linear space (a vector space with an extra bracket product).

\m 

\n{\bf Remark 2} On the other hand, remarkably little information is lost in passing from a Lie group to the corresponding Lie algebra.  
For instance, the representations of $\Frg$ determine those of $\lFrg$.

\m 

\n{\bf Remark 3} That the Lie bracket arises from considering Lie group structure in the vicinity of the identity can be seen for instance from 
restricting thereto the differentiation of conjugation, 
\be 
\left. \frac{\d}{\d \nu} \left( g H g^{-1} \right) \right|_{\nu = 0}  \es  \mbox{\bf [} \, G \mbox{\bf ,} \, H \, \mbox{\bf ]} \m , 
\ee 
for $G = g^{\prime}(0)$ and using $g(0) = 1$.

\subsection{From Lie algebras to Lie groups$^*$}

\n{\bf Structure 1} Working in the opposite direction, the globalizing move is 
\be 
\mbox{solving \m exp($X$)exp($Y$) = exp($Z$) \m for \m $Z$ \m when \m $X$, $Y$ \m do not necessarily commute} \m . 
\ee
To connect with the literature, this is often attributed to Baker, Campbell and Hausdorff.
Hamilton \cite{Hamilton} already intuited the first correction to be 
\be 
\half\mbox{\bf [} \, X \mbox{\bf ,} \, Y \, \mbox{\bf ]}  \m .
\ee 
Schur \cite{Schur} worked on subsequent correction terms.  
One issue is these terms' explicit formulae.
Another is proving that these depend on $X$ and $Y$ through successive uses of commutators alone: 
\be
Z = Z( \, [ \m , \, \m ] \m \mbox{ alone} \, )  \m .  
\ee
It was Hausdorff \cite{H06} who first produced a complete proof of this. 
Dynkin \cite{Dynkin} subsequently tidied up by producing a closed-form expression for the general-$n$ term manifestly in terms of commutators alone.  
By this, let us use `Hamilton--Schur--Hausdorff--Dynkin' formulae and `Hausdorff's Lie-Globalization Theorem'.  
For junior readers who would like a proof they can readily follow, 
Stillwell's account \cite{Stillwell-Lie} of Eichler's proof \cite{E68} -- induction using just basic algebra -- is recommended. 

\m 

\n{\bf Remark 1} Hausdorff's Lie-globalization Theorem signifies that the global {\it group commutator}  
\be 
g \, h \, g^{-1} \, h^{-1}
\ee
for a Lie group is totally controlled by the local Lie algebra's commutator (\ref{Lie-Bra}). 
This is how the above-mentioned remarkably little loss of information comes about.  

\m 

\n{\bf Structure 2} The Jacobi identity carries over to the {\it Hall--Witt} alias {\it three subgroups identity} of Lie groups \cite{Serre-Lie}
\be 
\mbox{\bf [} \, g \mbox{\bf ,} \, h^{-1} \mbox{\bf ,} \, k \, \mbox{\bf ]}^h \times \mbox{(cycles)}  \es  1 \m ,
\ee 
where the exponent $h$ denotes conjugation by $h$.  

\m 

\n{\bf Remark 3} Only the presence of other sectors supported by discrete transformations, giving further components not connected to the identity, 
is lost in the passage from Lie groups to Lie algebras, and then not remembered in the reverse passage. 
This amounts to a small amount of information of a topological nature.

\subsection{Some basic examples of Lie groups and Lie algebras}\label{Basic-Ex}

\n{\bf Examples 1} [{\it of Lie groups}]. The easier part of Lie Theory consists of the $SO(d)$, $O(d)$, $SU(d)$, $U(d)$ and $Sp(d)$ series: 
(special) orthogonal, (special) unitary and symplectic groups.
`Special' here means unit-determinant, with the symplectic case being already-special. 

\m 

\n{\bf Remark 1} These three special series are moreover connected, with $SO(d)$ being the component connected to the identity of $O(d)$, and $SU(d)$ likewise for $U(d)$.  
These are continuous groups, whereas $O(d)$ and $U(d)$ are mixed, i.e.\ containing further discrete transformations: reflections.  

\m 

\n{\bf Remark 2} All five of these series can moreover be viewed as matrix groups \cite{Stillwell-Lie}, 
a position \cite{V29} that greatly simplifies the ensuing Lie Theory. 

\m 

\n{\bf Examples 2} [{\it of Lie algebras}]. Corresponding Lie algebras: $so(d)$, $su(d)$, $sp(d)$.  
These three special series are probably most readily picked out \cite{C46, Stillwell-Lie} 
by considering isometry groups in $\mathbb{R}^n$, $\mathbb{C}^n$ and $\mathbb{H}^n$ (quaternionic space). 
The explanation of why there are only three such series is thereby rooted in Algebra, 
with the more limited octonians providing a route to envisage the remaining handful of exceptional Lie algebras \cite{Cartan55, FHBook}.
%

\subsection{Contractions of Lie algebras}

A {\it contraction} is a limiting operation on a Lie algebra parameter \cite{Gilmore}, arising as a singular limit 
in a change of basis, $L(\epsilon)$.
More specifically, when it exists, 
\be 
\s{\mbox{lim}}{\mbox{\scriptsize{$\epsilon \longrightarrow 0$}}} 
\left(
L^{-1}(\epsilon) \mbox{\bf |[} \, L(\epsilon) g \mbox{\bf ,} \, L(\epsilon) \, h \, \mbox{\bf ]|} 
\right)
\ee
is capable of being distinct from $\mbox{\bf |[} \, g \mbox{\bf ,} \, h \, \mbox{\bf ]|}$, 
by the singular basis change altering some of the structure constants. 
We require this for Article IX.  

\section{Generalized Killing equations}\label{GKEs}

\n{\bf Structure 1} The Lie derivatives with respect to the totality of vector fields 
%
%
on a manifold form an infinite-dimensional Lie algebra of infinitesimal diffeomorphisms, 
\be 
diff(\Frm)  \m .   
\label{diff}
\ee 
\n{\bf Structure 2} 
\be 
\phi_*\bG  =  \bG
\ee 
defines a {\it symmetry} for the general geometrical object $\bG$.
This means that $\bG$ is invariant under displacements along the integral curves of the corresponding vector field.  
Such vector fields enter by restricting (\ref{diff}) to those that solve the corresponding version of the following equation. 

\m 

\n{\bf Structure 3} If one applies an infinitesimal transformation 
\be 
\bx \m \longrightarrow \m \bx^{\prime}  \es  \bx + \epsilon \, \bX
\ee 
to a particular $\bG$'s transformation law, 
\be 
\pounds_{\sbX} \bG = 0 
\label{LG}
\ee 
arises to first order \cite{Eisenhart33, Yano55}.  

\m 

\n{\bf Structure 4} Some tensors and other geometrical objects moreover have further significance as 
{\it further levels of geometrical structure}, $\bsigma$.   
Examples include \cite{Yano55, KN1, Yano70, Kobayashi} the metric tensor $\bg$ and the connection $\bgamma$, 
as well as simlarity and conformal equivalence classes of metrics, and affine and projective equivalence classes of connections.

\m 

\n{\bf Structure 5} Let us use the notation 
\be
\langle \, \Frm, \bsigma \, \rangle
\ee
for a differentiable manifold $\Frm$ equipped with geometrical level of structure $\bsigma$.  

\m 

\n{\bf Structure 6} Such cases' version of (\ref{LG}), 
\be 
\pounds_{\sbX} \bsigma = 0  \m , 
\ee 
is known furthermore as a {\it generalized Killing equation} \cite{Yano55, Yano70}.  

\m 

\n{\bf Structure 7} The solutions thereof are {\it generalized Killing vectors} $\bupxi = \bupxi(\bx)$ 
(for $\bupxi$ the covector corresponding to $\bX$ \cite{Yano70}).

\m 

\n{\bf Structure 8} For a particular $\langle \, \Frm, \bsigma \, \rangle$ these moreover close \cite{Kobayashi} as a Lie algebra, 
\be
\mbox{\bf [} \, \uc{\bupxi}(\bx) \mbox{\bf ,} \, \uc{\bupxi}(\bx) \, \mbox{\bf ]}  \es  \uc{\uc{\uc{\biZ}}} \, \uc{\bupxi}(\bx)           \m , 
\label{xi-Lie}
\ee
where $A, B, C$ are multi-indexes comprising both the corresponding spatial index $a, b, c$ 
and the generator-basis index $g$, and $\biZ$ are the corresponding {\it structure constants}.
As a Lie algebra, this corresponds to the continuous connected component of the identity part of the automorphism group,  
\be 
Aut(\Frm, \bsigma)  \m . 
\label{Aut}
\ee 
Hence we denote this by 
\be 
aut(\Frm, \bsigma)  \m .
\ee 
\n{\bf Remark 1} Some insightful true-names are {\it continuous automorphism group finding equation} in place of generalized Killing equation, 
and {\it continuous automorphism generators} in place of generalized Killing vectors.  

\m

\n{\bf Remark 2} For use in Sec \ref{LIToI}, let us give firstly that the infinitesimal-generator form of automorphism vectors is 
\be
\uc{\bupxi}  \es  \u{\uc{\biG}}(\bfB) \cdot \u{\nabla}_{\sbfB}    \m . 
\ee
Secondly, that the (\ref{Aut}) are finite \cite{Eisenhart33, Yano55} subalgebras of the (\ref{diff}), 
in the sense of being a finite count of independent generators.

\subsection{Example 1) Killing vectors and isometries}

\n{\bf Structure 1} {\it Isometries} (in the geometrical, rather than metric space context) are $\Frm$-diffeomorphisms that additionally preserve the metric structure $\bm$.   
This is additionally the 
\be
\phi^*\bm  =  \bm
\ee 
subcase of the aforementioned more general definition  of symmetries for objects $\bG$.

\m 

\n{\bf Structure 2} Isometries take the infinitesimal form  
\beq
\epsilon_{AB} \m \longrightarrow \m \epsilon_{AB} - 2 \, \cD_{(A}\xi_{B)}  \m 
\eeq 
for $\cD_A$ the covariant derivative corresponding to $\Frm$.  
It may be useful to some readers to note that this exhibits some parallels with the transformations of Gauge Theory,   
\be 
\mA_{\sfA} \m \longrightarrow \m \mA_{\sfA} - \pa_{\sfA}\Lambda  \m .   
\ee 
\n{\bf Structure 3} Isometries are found by solving the {\it Killing equation}: the first equality in 
\beq
0  \es   \pounds_{\xi} \mm_{AB} 
   \es   2 \, \cD_{(A} \xi_{B)} 
   \=:  (\cK \xi)_{AB}                \m . 
\label{Killing}
\eeq
The second equality here is a simple computation, whereas the final definition is for {\it Killing form}     $(\cK \,\xi)_{AB}$ 
                                                                                   or {\it Killing operator} $\cK$.  
(\ref{Killing})'s solutions are the {\it Killing vectors} of $\langle \, \Frm, \, \bm \, \rangle$.

\m 

\n{\bf Structure 4} Isometries form the familiar isometry group $Isom(\Frm) = Aut(\Frm, \bsigma)$ subcase of automorphism group. 

\m 

\n{\bf Killing's Lemma} \cite{Yano55} is that 
\beq
\cD_{A}\cD_{B}\xi_{C}  \es  - {\cR_{BCA}}^{D}\xi_{D} \m ,
\label{Killing-Lemma}  
\eeq
for ${\cR_{BCA}}^{D}$ the Riemann curvature tensor of $\langle \, \Frm, \, \bm \, \rangle$.

\subsection{Example 2: similarity counterpart}\label{CGKEs}

\n{\bf Structure 1} A {\sl similarity} (sometimes alias {\it homothety}) \cite{MacCallum} can now be understood to be a diffeomorphism 
that additionally preserves the metric structure up to constrant rescaling, 
\be 
\phi^*\bm  =  c^2 \, \bm                                              \m .
\ee  
This is also the $\bT = \bm$ subcase of the more general definition 
\be 
\phi^* \bT  =  c^{2 \, w} \bT
\ee 
of a {\it similarity with weight w} of a tensor $\bT$. 

\m 

\n{\bf Structure 2} The infinitesimal form taken by a rescaling  transformation is 
\beq
\epsilon_{AB} \longrightarrow \epsilon_{AB} + c^2 \mm_{AB}            \m .
\eeq 
\n{\bf Structure 3} Similarities are found by solving the {\it similarity Killing equation}: the first equality in  
\beq  
2 \, c \, \mm_{AB}  \m = \m \pounds_{\xi}\mm_{AB} = 2 \, \cD_{(A} \xi_{B)}                          \m , 
\label{SKE}
\eeq
also written schematically as 
\be 
(\cS \xi)_{AB} = 0 
\ee 
for {\it similarity Killing form}     $(\cS \,\xi)_{AB}$ or 
    {\it similarity Killing operator} $\cS$.  
(\ref{SKE})'s solutions are the {\it similarity Killing vectors} of $\langle \, \Frm, \bar{\bm} \, \rangle$ for $\bar{\bm}$ the metric structure up to constant rescaling.

\m 

\n{\bf Structure 4} Similarities form the {\it similarity group} $Sim(\Frm) = Aut(\Frm, \bm)$ subcase of automorphism group. 
We leave their analogue of Killing's Lemma as a simple exercise for the reader.

\m 

\n{\bf Remark 1} See e.g. \cite{A-Killing} for more about these examples and an outline of their conformal, affine and projective counterparts, 
or \cite{Yano70, Yano55} for their detailed theory.

\subsection{Further theory of (generalized) Killing equations}

\n{\bf Remark 1} Generalized Killing equations are {\it homogeneous linear first-order systems} of PDEs. 

\m 

\n{\bf Remark 2} These are in general {\it over-determined}, lending themselves to having a lack of nontrivial solutions. 
Trivial solutions -- the zero, or in some cases constant, vectors -- are guaranteed by homogeneity. 
However, only nontrivial solutions count as Killing vectors: a nontrivial kernel condition.     

\m 

\n{\bf Remark 3} Whether over-determined PDE systems admit (nontrivial) solutions is down to whether they satisfy {\it integrability conditions}. 
For instance, Killing's Lemma (\ref{Killing-Lemma}) can furthermore be interpreted as an integrability condition for Killing's equation to be solvable. 

\m

\n{\bf Remark 4} The generic $\langle \,  \Frm, \bsigma \, \rangle$ admits no notrivial generalized Killing vectors.  
Many nontrivialities moreover require at least two Killing vectors to be present (and non-commuting at that) \cite{MacCallum}; 
these are {\sl fairly highly} nongeneric manifold geometries $\langle \, \Frm , \bsigma \, \rangle$.  
Each kind of generalized Killing equation has moreover a manifold-dimension-dependent maximal number of independent generalized Killing vectors \cite{Yano55}; 
this is the most special, i.e.\ least generic case. 
For Killing's equation itself, these are the maximally-symmetric spaces, which are required to be of constant curvature 
(so e.g.\ $\mathbb{R}^n$ and $\mathbb{S}^n$ are such). 

\m 

\n{\bf Remark 5} At least in all the cases mentioned above, the generalized Killing equation is moreover elliptic (in the basic sense to be found in e.g.\ \cite{John}). 

\m 

\n{\bf Remark 6} Solving the generalized Killing equation is Sec \ref{FOP} backward route; 
this is quite generally technically harder than the corresponding forward route.  
This is moreover on two counts: aside from involving integration rather than differentiation, 
(generalized) Killing equations' ellipticity renders them globally sensitive, whereas Lie's elimination is just a local affair.

\section{Relationalism as implemented by Lie derivatives}\label{Rel-Lie-Deriv}

\subsection{Temporal Relationalism implementing (TRi) Lie derivatives}

\n{\bf Remark 1} Passing from direction-of-derivation vector $\d \bfQ/\d \lambda$ to the TRi $\d \bfQ$ does not alter Lie derivative status.  

\m 

\n{\bf Remark 2} Ensuing reinterpretation of tangent bundles as configuration-change, rather than configuration-velocity, 
bundles remains within Sec \ref{VT}'s remit.

\subsection{Configurational Relationalism correcting Lie derivatives}

\n{\bf Remark 1} For $\lFrg$ acting on $\langle \, \FrQ , \, \bsigma \rangle$, using Article II's nomenclature, 
we first perform an `Act' move: $\bfO \m \longrightarrow \m \s{\rightarrow}{\lFrg}_g \bfO$ 
that is locally implementable by $\pounds_g \bfO$.  

\m 

\n We then follow this up with an `All' move, such as integration, averaging or extremization for $g$ over $\lFrg$.  

\m 

\n We can, more generally, interpose a $Maps$ move between these two moves.  

\m 

\n{\bf Remark 2} This works just as well for the TRi $\pounds_{\d g}\bfO$ with cyclic differentials $\d g$ providing direction-of-derivation. 
Thereby, Articles V and VI fit our Lie-theoretic rubric as well as Article II does.  

\m 

\n{\bf Remark 3} Ensuing mixed tangent--cotangent bundles also remain within Sec \ref{VT}'s remit.    

\m 

\n{\bf Remark 4} Thus both TRi and its combination with CRi to form Ri -- Relationalism implementation -- carry over.

\subsection{Spacetime Relationalism correcting Lie derivatives}

\n This parallels the previous subsection, under the substitutions $\lFrg_{\sS}$ for $\lFrg$ and 
$\langle \,  \Spacetime, \, \bsigma \, \rangle$ for $\langle \, \FrQ , \, \bsigma \rangle$, as per Article X, 
and there now being no call for combination with Temporal Relationalism.

\section{Further differentiable structures$^*$}\label{+Diff}

This permits further global considerations for the Problem of Time and Background Independence.

\subsection{The simpler state spaces}

{\bf Definition 1} {\it Constellation space} 
\be
\FrQ(\Frm^d, N) \es \bigtimes_{i = 1}^n \Frm^d   \m . 
\ee 
\n This models $N$-Body Problems in Classical Mechanics.

\subsection{Stratification}

\n{\bf Remark 1} At the level of configuration space $\FrQ$, 
independent generators $\scg \in \lFrg$ remove one degree of freedom each. 

\m 

\n On the other hand, in single bundle spaces such as $\FrT(\FrQ)$ and $\Phase$, 
generators can use up 2, or just 1, degree of freedom.   
There is often, though not always, a relation between using up 2 degrees of freedom and being gauge. 
%

\m 

\n At the level of the space of spacetimes $\Spacetime$, 
independent generators $\scg \in \lFrg_{\sS}$ remove one degree of freedom each. 

\m 

\n{\bf Structure 1} The quotient space -- {\it reduced configuration space} -- 
\be 
\w{\FrQ}  \:=  \frac{\FrQ}{\lFrg}
\ee 
is in general {\it stratifed} \cite{W65, T69}: it can vary in dimension from place to place, and thus not be locally Euclidean and thus not be a manifold. 
This is a more general topological space than a topological manifold, with {\it singular} differential structure thereupon.   
It can still however be modelled locally piece by piece using standard differential structure.
Examples include $N$-body Problem reduced configuration spaces (Articles II and V), and GR's superspace (Article VI). 

\m 

\n{\bf Remark 1} {\it Singular varieties} \cite{Hartshorne} and {\it catastrophes} \cite{Arnold} are better-known cases of singular differential structures.   
There is plentiful evidence that continuum modelling, if carried out in sufficient detail 
(in particular as regards eliminating redundancies), universally gives singular differentiable spaces  
as 'the next approximation after topological-and-differentiable manifolds'.

\m 

\n{\bf Remark 2} In stratified manifolds, some generators may then be inactive on some strata. 
E.g. only an $SO(2)$ subgroup of $SO(3)$ is active on the 3-$d$ 3-body problem's collinear stratum.
This kind of phenomenon being a global matter, we do not return to it in this Series; 
see \cite{A-Monopoles, Minimal-N, A-Killing, A-Cpct} for further details.  

\m 

\n{\bf Remark 3} Single bundle arenas formed by quotienting by a Lie group, in particular {\it reduced phase space} 
\be 
\w{\Phase} \:= \frac{\Phase}{\lFrg} \m ,
\ee 
can also exhibit stratifaction \cite{PflaumBook}.   
So can the {\it reduced space of spacetimes}
\be 
\w{\Spacetime}  \es  \frac{\Spacetime}{\lFrg_{\sS}}  \m , 
\ee 
for instance GR's superspacetime (Article X).  

\m 

\n{\bf Remark 4} The more straightforward stratified spaces arising from such reductions take after the next subsection, 
whereas the final subsection indicates  substantial further difficulties with the less straightforward cases.

\subsection{LCHS spaces}

\n{\bf Definition 1} A {\it LCHS} space is a topological space that is locally-compact, Hausdorff and second-countable.  

\m 

\n{\bf Example 1} Manifolds are a subcase: 
\be 
\mbox{LEHS } \Rightarrow \mbox{ LCHS}   \m . 
\ee
{\bf Proposition 1} LCHS spaces are moreover more general, and remain tractable along the following lines \cite{Munkres, Lee2}. 
\be 
\mbox{LCHS} \m \Rightarrow \m \mbox{P} \m \mbox{ (paracompactness) } .  
\ee 
{\bf Remark 1} In turn, this permits LCHS stratified manifolds with the following features. 

\m 

\n 1) partitions of unity and thus a practicable notion of {\it integration}. 

\m 

\n 2) The {\it Shrinking Lemma} (originally due to Lefschetz \cite{L42}, see e.g.\ \cite{Munkres, Lee2} for a modern presentation) 
as a further guarantee of, and tool for, local treatment. 

\m 

\n 3) Inclusion of some {\it singular manifolds}, including some of the better-behaved {\it stratified manifolds} \cite{W46, T55, W65, T69}. 

\m 

\n 4) Fibre bundles do not suffice in treating stratified manifolds. 
General bundles, presheaves, or the more mathematically powerful sheaves fill this role instead. 
In the LCHS case, there is moreover collapse from more general Sheaf Methods \cite{Wells, Wedhorn, Bredon} and Sheaf    Cohomology \cite{Wedhorn, Hartshorne, Bredon, Iversen} 
                      to just \v{C}ech Methods                            and \v{C}ech Cohomology \cite{HY, BT82}.  

\m 

\n {\it Kreck's stratifolds} \cite{Kreck} are an example of nicely controllable (LCHS stratified manifold, sheaf) pairs.  
%

\m 

\n Alternative differential spaces approaches, were adapted for use in considering nice stratified manifolds by \'{S}niatycki \cite{SniBook}. 

\m 

\n{\bf Remark 2} Shrinking one labelled domain to another moreover gives an example of how it has become useful to label domains and neighbourhoods.  

\m 

\n{\bf Remark 3} Examining domains and neighbourhoods in detail enables various other 20th century topological tools as well, 
following from establishing e.g.\ that one has a {\it tubular neighbourhood} \cite{GP} or a {\it collar neighbourhood} \cite{Lee2}.    
All in all, Lie Theory rooted in detailed topology is argued to be a major improvement on the original Lie Theory. 

\m 

\n{\bf Remark 5} All in all, locality lets one keep on using Lie's Mathematics in certain regions herein.  
Killing's Mathematics, however, has global flavour due to entailing solution of elliptic PDEs.

\m 

\n{\bf Remark 6} While moment maps, reduction... usually assume the quotient will be a manifold, 
less well-known and less used singular counterparts have been developed (e.g.\ \cite{Landsman} contains a review). 
See also e.g.\ \cite{PflaumBook} for an outline of stratified phase spaces.

\subsection{Less structured stratified spaces}

\n{\bf Remark 1} Affine \cite{PE16} and Projective reduced configuration spaces exhibit the following major complication.

\m 

\n{\bf Definition 1} A topological space $\FrX$ is {\it Kolmogorov} \cite{Willard, Engelking} 
if whichever 2 distinct points $x \neq y$ in $\FrX$ are {\it topologically distinguishable}, meaning that  
\be 
\mbox{$\exists$ \m \m at least one open set \m $\FrU$ \m  such that \m $x \in \FrU$ \m but \m $y \not{\hspace{-0.05in}\in} \m \FrU$}  \mma  
                                                 \mbox{or } \m  \mbox{$x \in \FrU$ \m  but \m $y \not{\hspace{-0.05in}\in} \m \FrU$}  \m .
\ee 
\n{\bf Remark 2} Kolmogorov separability is moreover much weaker than, and {\sl qualitatively distinct} from, Hausdorff separability; 
essentially no Analysis is now supported. 

\m 

\n{\bf Structure 1}  Quotienting a manifold (in particular for us a state space) by a Lie group does not in general preserve Hausdorffness. 

\m 

\n Some examples of such state space quotients are moreover merely-Kolmogorov in separability \cite{A-Cpct}.  

\m 

\n{\bf Definition 2} A group action of $\lFrg$ on a topological space $\FrX$ is \cite{Lee1}  
\be 
\mbox{{\it proper} if each compact subset } \m  \FrK \, \subseteq \, \FrX \m \mbox{ has action inverse-image } \m \Phi^{-1}(\FrK) \m \mbox{ itself be compact} \m . 
\ee
\n{\bf Remark 3} Proper actions guarantee Hausdorff separability is maintained under quotienting \cite{A-Cpct}; 
this guarantee is clearly absent for the above-mentioned affine and projective cases, as well as for further quotients of Minkowski spacetime.

\section{Lie Algorithms}\label{LA}

\subsection{Clebsch versus Lie}

\n{\bf Position 1} Clebsch \cite{Clebsch} considers the case in which all brackets of generators are already contained among linear combinations of generators; 
for constant coefficients, this returns (\ref{Str-Const}). 

\m 

\n{\bf Position 2} Lie \cite{Lie} however emphasizes the significance of extending this to the case in 
which new generators $\sch$ can be discovered amongst the brackets of already-known generators $\scg$.

\m 

\n He also points out that this in general requires proceeding recursively, so 
\be
\mbox{\bf |[} \, \scg \mbox{\bf ,} \, \sch \, \mbox{\bf ]|}  \mma  
\mbox{\bf |[} \, \sch \mbox{\bf ,} \, \sch \, \mbox{\bf ]|} 
\ee
now need to be investigated, and might produce further generators $\sch_{(2)}$ 
(labelling $\scg$ as $\sch_{(0)}$ and $\sch$ as $\sch_{(1)}$; we call $\scg$, $\sch$ together, i.e.\ $\sch_{(0)}$, $\sch_{(1)}$ together, $\scg_{(1)}$, and so on).  

\m 

\n He finally envisages the possibility of a trivial outcome, by all degrees of freedom getting used up.  

\m 

\n{\bf Remark 1} This actual procedure of Lie's contains half of the cases in what this Series terms `Little Lie Algorithm'. 
The other half follows from observing and generalizing one insight and one application of Dirac's, as follows.

\subsection{Dirac's insight generalized}

\n{\bf Remark 1} We additionally capitalize on Dirac's envisaging the need to allow for the possibility of inconsistency.  
This allows for an algorithm with {\sl selection principle} properties.  

\m 

\n{\bf Example 1} My considering what happens \cite{AMech, ABook} in attempting to jointly include the special-conformal generator and the affine generator 
in the Foundations of Geometry setting vindicates this as a general Lie, rather than just Dirac, feature. 
Namely, in flat space, one has to chose {\sl one of} the special conformal generator or the affine generator (see Sec VII.6.2).

\subsection{Lie-appending as a generalization of Dirac-appending}

\n{\bf Remark 1} Suppose we are dealing with an application of Lie's Algorithm that comes with a procedure for appending by auxiliaries. 

\m 

\n{\bf Example 1} The most well-known case of this is Dirac-appending in his generalized Hamiltonian treatment of constraints \cite{Dirac51, Dirac58, Dirac}. 
Here `bare' Hamiltonians have constraints appended by Lagrange multipliers to form some kind of notion of extended Hamiltonian 
(see \cite{Dirac, HTBook} and Article III).

\m 

\n{\bf Structure 1} Then relations amongst some of these auxiliaries -- {\it specifier equations} -- 
may arise from the iterations of taking Lie brackets of existing objects.

\subsection{Lie- alias generator-weak generalization of Dirac- alias constraint-weak equality}

{\bf Definition 1} Let us use 
\be 
\approx
\ee 
to mean equality up to a linear function(al) of generators: {\it Lie-} alias {\it generator-weak equality}.   
This extends Dirac's use of the same symbol to mean equality up to a linear function(al) of constraints: {\it Dirac-} alias {\it constraint-weak equality} 

\m 

\n{\bf Remark 1} In contrast, strong equality 
\be 
= 
\ee 
is just equality in the usual sense; this clearly does not require any `constraint' or `generator', or `Dirac' or `Lie', qualifications.  

\m 

\n{\bf Definition 2} Let us finally introduce  
\be 
\peq
\ee 
to denote {\it portmanteau equality}: strong or weak. 
Having already used this for {\it Dirac-} alias {\it constraint-portmanteau equality} in Article VII, 
we now extend it to mean     {\it Lie-}   alias {\it  generator-portmanteau equality}.

\subsection{Lie's Little Algorithm}

\n{\bf Definition 1} {\bf Lie's Little Algorithm} \cite{Dirac} consists of evaluating Lie brackets between a given input set of generators 
so as to determine whether these are consistent and complete.  
Three possible types of outcome are allowed in this setting.    

\m 

\n{\bf Type 0)} {\bf Inconsistencies}.

\m 

\n{\bf Definition 2} Let us refer to equations of all other types arising from Lie Algorithms as {\it ab initio consistent}.

\m 

\n{\bf Type 1)} {\bf Mere identities} -- equations that reduce to 
\be 
0 \peq 0                       \m .  
\ee 
\n{\bf Type 2)} {\bf New generators}, e.g.\ discovering rotations and scalings as mutual integrabilities of translations and special-conformal transformations 
(see Sec IX.9).  

\m 

\n{\bf Definition 3} The {\bf Extended Lie's Little Algorithm} is for cases that come with an appending procedure.  
In this case, an additional type of equation can arise, as follows.

\m 

\n{\bf Type 3)} {\bf Specifier equations}.   

\m 

\n{\bf Remark 1} Dirac's Little Algorithm, and the TRi-Dirac Little Algorithm, are examples of Extended Lie's Little Algorithms.  

\m 

\n{\bf Definition 4} With type 1)'s mere identities having no new content, let us call types 2) to 4) `{\it nontrivial ab initio consistent equations}'.  
Note that we say `equations', not `generators', to include type 4)'s specifier equations; 
in cases with no such, we could as well in the Little Algorithm say `generators'. 

\m 

\n{\bf Remark 2} If type 2) occurs, the resultant objects are fed into the subsequent iteration of the algorithm, 
which starts with the extended set of objects.  
One is to proceed thus recursively until one of the following termination conditions is attained. 

\m 

\n{\bf Termination Condition 0) Immediate inconsistency} due to at least one inconsistent equation arising.

\m 

\n{\bf Termination Condition 1) Combinatorially critical cascade}. This is due to the iterations of the Lie Algorithm producing a cascade of new objects 
down to the `point on the surface of the bottom pool' that leaves the candidate with no degrees of freedom.   
I.e.\ a combinatorial triviality condition.    

\m 

\n{\bf Termination Condition 2) Sufficient cascade}, i.e.\ running `past the surface of the bottom pool' of no degrees of freedom 
into the `depths of inconsistency underneath'.

\m 

\n{\bf Termination Condition 3) Completion} is that the latest iteration of the Lie Algorithm has produced no new nontrivial consistent equations, 
indicating that all of these have been found. 

\m 

\n{\bf Remark 3} Our input candidate set of generators is either itself {\it complete} -- Clebsch -- 
                                                     or {\it incomplete} -- `nontrivially Lie' -- 
													 depending on whether it does not or does imply any further nontrivial objects.
If it is incomplete, it may happen that Lie's Algorithm provides a completion, by only an {\it combinatorially insufficient cascade} arising, 
from the point of view of killing off the candidate.  
														 
\m 														 

\n{\bf Remark 4} So, on the left point of the trident, Termination Condition 3) is a matter of acceptance 
of an initial candidate set of generators alongside the cascade of further objects emanating from it by Lie's Algorithm.   
(This acceptance is the point of view of consistent closure; further selection criteria might apply.)    			
This amounts to succeeding in finding -- and demonstrating -- a `Lie completion' of the initial candidate set of generators.   											 

\m  

\n{\bf Remark 5} On the right point of the trident, Termination Conditions 0) and 2) are a matter of rejection of an initial candidate set of generators.  
The possibility of either of these applying at some subsequent iteration justifies our opening conception in terms of ab initio consistency.

\m 

\n{\bf Remark 6} On the final middle point of the trident, Termination Condition 1) is the critical case on the edge between acceptance and rejection; 
further modelling details may be needed to adjudicate this case.

\m 

\n{\bf Remark 7} In each case, {\it functional independence} \cite{Lie} is factored into the count made; 
the qualification `combinatorial' indicates Combinatorics not always sufficing in having a final say.  
For instance, Field Theory with no local degrees of freedom can still possess nontrivial global degrees of freedom. 
Or Relationalism can shift the actual critical count away from zero, e.g.\ by requiring a minimum of two degrees of freedom 
so that one can be considered as a function of the other.  
If this is in play, we use the adjective `relational' in place of (or alongside) `combinatorial'.  

\m 

\n{\bf Remark 8} In the general Lie context, we note that the term `prolongation' has quite widespread use \cite{Olver}. 
We prefer `cascade' for our particular use, however, from the point of view of emphasizing the possibility of multiple iterations, 
including `all the way to the bottom pool'.  

\m 

\n{\bf Remark 9} In detailed considerations, clarity is often improved by labelling each iteration's set of objects by the number of that iteration.

\subsection{Classness and rebracketing generalized from Dirac to Lie}

\n{\bf Definition 1} {\it Lie-first-class objects}  
\be 
\bscf \m \mbox{ indexed by } \m \ff
\ee  
are those that close among themselves under Lie brackets.  

\m 

\n{\bf Definition 2} {\it Lie-second-class objects} \cite{Dirac, HTBook} 
\be 
\bscs\bsce \m \mbox{ indexed by } \m  \fe
\ee 
are those that are not Lie-first-class (a definition by exclusion).  

\m 

\n{\bf Diagnostic} For the purpose of counting degrees of freedom, in single bundles
Lie-first-class objects use up two each whereas Lie-second-class objects use up only one.  

\m 

\n Purely on base spaces, including configuration spaces, however, only first-classness is possible, 
with all independent Lie objects using up one degree of freedom each.
This extends to `half-polarizations': a general setting for QM.  

\m 

\n{\bf Remark 1} Lie-first-class objects are generators, but are not necessarily gauge generators; 
for now we give the canonical example of Dirac's Conjecture \cite{Dirac} failing \cite{HTBook, VII} as a counterexample.   

\m 

\n{\bf Remark 2} We are now to envisage the possibility of Lie-second-class objects arising at some iteration in the Lie Algorithm.

\m 

\n{\bf Remark 3} Lie-second-class objects $\Sec$ can moreover be slippery to pin down.
This is because Lie-second-classness is not invariant under taking linear combinations of Lie objects. 
Linear Algebra dictates that the invariant concept is, rather, {\sl irreducibly Lie-second-class objects}   
\be 
\bsci \m \mbox{ indexed by } \m \fii  \m .
\ee  
\n{\bf Proposition 1} Irreducibly Lie-second-class objects can be factored in by replacing the incipent Lie bracket 
                                                                                        with the {\it `Lie--Dirac bracket'} 
\beq
\mbox{\bf |[} \,     F    \mbox{\bf ,} \, G    \, \mbox{\bf ]|}  \mbox{}^{\mbox{\bf *}}  :=  \mbox{\bf |[} \,    F   \mbox{\bf ,} \, G     \, \mbox{\bf ]|} - 
                                                                                      \mbox{\bf |[} \,    F   \mbox{\bf ,} \, \uc{\bsci} \, \mbox{\bf ]|}      \cdot  
                                                                                      \mbox{\bf |[} \,  \uc{\bsci} \mbox{\bf ,} \, \uc{\bsci} \, \mbox{\bf ]|}^{-1} \cdot
                                                                                      \mbox{\bf |[} \,  \uc{\bsci} \mbox{\bf ,} \, G     \, \mbox{\bf ]|}             \m . 
\eeq
Here the --1 denotes the inverse of the given matrix, and each $\cdot$ contracts the objects immediately adjacent to it.   

\m 

\n{\bf Remark 4} The role of brackets initially played by the Lie brackets is thus in general taken over by the Lie--Dirac brackets.
This extends the way in which Dirac brackets take over the role of Poisson brackets. 
Dirac brackets can moreover be viewed geometrically \cite{Sni} as more reduced spaces' incarnations of Poisson brackets, and thus are themselves Poisson brackets. 
We thus point to the possibility that Lie--Dirac brackets can be viewed geometrically as more reduced spaces' incarnations of Lie brackets 
(though we have not in general proven this).    

\m 

\n{\bf Remark 5} Passage to Lie--Dirac brackets can moreover recur if subsequent iterations of the Lie Algorithm unveil further Lie-second-class objects.  

\m 

\n This leads to a notion of {\it final Lie--Dirac bracket}, meaning the {\it maximal Lie--Dirac bracket} by which all the Lie-second-class objects  
a candidate algebraic structure can produce have been factored out.

\subsection{Full Lie Algorithm}

\n Proceed as before, except that whenever second-class constraints appear, switch to (further) Lie-Dirac brackets that factor these in.  
This amounts to a fourth type of equation being possible, as follows. 

\m 

\n{\bf Type 4)} {\bf Further Lie-second-classness} may arise.

\m 

\n On the one hand, these could be {\it Lie-self-second-class}, meaning that brackets between new objects do not close. 

\m 

\n On the other hand, these could also be {\it Lie-mutually-second-class}, 
meaning that some bracket between a new object and a previously found object does not close. 
By which, this previously prescribed or found object was just {\it hitherto} Lie-first-class.
I.e.\ Lie-first-classness of a given constraint can be lost whenever a new constraint is discovered.  

\m 

\n{\bf Remark 1} This is as far as Dirac gets; subsequent discoveries in practice dictate the addition of a sixth type. 
Dirac knew about this \cite{Dirac}, commenting on needing to be lucky to avoid this at the quantum level. 
But no counterpart of it enters his classical-level Algorithm.  

\m 

\n{\bf Type 5} {\bf Discovery of a topological obstruction} to having a Lie algebraic structure of constraints.
The most obvious examples of this are anomalies at the quantum level; it is however a general brackets phenomenon rather than specifically a quantum phenomenon.  

\m 

\n Two distinct strategies for dealing with this are as follows. 

\m 

\n{\bf Strategy 1} Set a cofactor of the topological term to zero when the modelling is permissible of this. 
In particular strongly vanishing cofactors allow for this at the cost (or discovery) of fixing some of the theory's hitherto free parameters.

\m 

\n{\bf Strategy 2} Abandon ship. 

\m 

\n{\bf Remark 2} In Lie's Little Algorithm, everything stated was under the aegis of all objects involved at any stage are Lie-first-class, 
                                                                                         and that no topological obstruction terms occur. 

\m

\n{\bf Remark 3} It is logically possible to have topological obstructions in the absence of appendings or the scope to rebracket; 
this describes the sigificant case of the Foundations of Geometry.  
We thus coin {\bf Topologically-adroit Lie's Little Algorithm} for this case.   

\m 

\n{\bf Remark 4} Overall `Little' thus differs in meaning in passing from Dirac to general Lie.
`Little Dirac' (Article III) refers to not needing to evoke Dirac brackets, 
whereas `Little Lie' refers collectively to both no appending and no need or scope to evoke Lie--Dirac brackets. 
In particular, these conditions apply to working on just a configuration space rather than some larger bundle thereover.
This reflects that, on the one hand, extending by appending is natural in Dirac's more specific setting, and so goes without saying there. 
On the other hand, this is not in general part of Lie's broader setting, thus requiring the qualification `Extended Little' when it is included.

\section{Resultant Lie algebraic structures}\label{LAS}

\subsection{Lie algebras}\label{Lie-Algebra-2}

{\bf Structure 1} The end product of a successful candidate brackets structure's passage through the Lie Algorithm is a {\it Lie algebraic structure}.
This consists solely of bona fide Lie-first-class generators closing under Lie (or more generally Lie--Dirac) brackets. 

\m 

\n Lie algebra closure is of the schematic form 
\beq
\mbox{\bf |[} \,  \scg \mbox{\bf ,} \,  \scg \, \mbox{\bf ]|}  \speq  0                                                                     \m ,
\label{F-F}
\eeq
which is a portmanteau for the strong version
\beq
\mbox{\bf |[} \,  \scg \mbox{\bf ,} \,  \scg \, \mbox{\bf ]|}    =  0                                                                      \m , 
\label{F-F-S}
\eeq
and the weak version: 
\beq
\mbox{\bf |[} \,  \uc{\scg}  \mbox{\bf ,} \,  \uc{\scg} \, \mbox{\bf ]|}  \es  \uc{\uc{\uc{\biG}}} \cdot \uc{\scg}                                  \m .
\label{F-F-W}
\eeq

\subsection{Lie algebroids}\label{Lie-Algebroid}

\n{\bf Structure 1} If the generators still close under Lie brackets
\be
\mbox{\bf |[} \,  \uc{\scg} \mbox{\bf ,} \, \uc{\scg}\mbox{}^{\prime} \mbox{\bf ]| }   \es  \uc{\uc{\uc{\biG}}}(\bfB, \bic) \, \uc{\scg}^{\prime\prime}    \m . 
\label{L-Algebroid}
\ee
but with {\it structure functions} $\biG(\biB, \bic)$ instead of structure constants, we have a {\it Lie algebroid}; 
the $\biB$ are base objects, 
whereas the $\bic$ are constants.   

\m 
 
\n{\bf Remark 1} Cartan first considered this possibility in 1904 \cite{Cartan55}; 
Rinehart gave a first modern treatment in 1963 \cite{R63}. 
See e.g.\ \cite{G06} for an introductory exposition and \cite{Algebroid1, Vaisman, CM, GS08} for further details.  
Algebroids can moreover be viewed as arising by observation in carrying out Lie algorithms.  
As further motivation, GR's Dirac algebroid (Sec III.2.19) is of this form, 
kinematical quantization's \cite{M63, I84} modern reformulation \cite{Landsman} in terms of Lie algebroids, 
and a fourth reason is given in Sec \ref{LAR}.

\m 

\n{\bf Remark 2} It may be possible to fix the constants $\bic$ to induce {\it strong avoidance of algebroidness}.  

\m 

\n{\bf Remark 3} Structure functionals $\biG(\bic; \bfB]$ are also possible. 
We are furthermore to include cases in which the $\biG$ are {\sl operator-valued}, 
be that differential-operator-valued in GR's Dirac algebroid $\biG(\bic, \nabla_{\sbfB}; \bfB]$, 
     or      quantum-operator-valued in kinematical quantization.   

\m 

\n{\bf Remark 4} Much as Lie algebras are the local version of a Lie group in the vicinity of the identity, 
Lie groupoids are for Lie algebroids. 
See e.g.\ \cite{Landsman} for more on Lie groupoids.

\subsection{Nontrivial $\Theta$ \m $^*$}\label{Theta}

\n{\bf Remark 1} A more general outcome is the inhomogenous-linear form  
\beq
\mbox{\bf |[} \,   \uc{\scg}  \mbox{\bf ,}  \,  \uc{\scg}^{\prime}  \, \mbox{\bf ]|}    \es  
\uc{\uc{\uc{\biG}}}[\bfB, \bic] \, \u{\scg}                                           \m + \m  
\uc{\uc{\uc{\biN}}}[\bfB, \bic] \, \u{\scg}^{\sn\se\sw}                               \m + \m  
  \uc{\uc{\bTheta}}[\bfB, \bic]                                                         \m ,   
\label{CO-Theta}
\eeq
where the extra linear terms $\scg^{\sn\se\sw}$ arise as integrabilities and the zeroth-order `obstruction terms' $\Theta$ can underly topological obstructions.   
Some possible outcomes in particular are as follows. 

\m 

\n 1) Perhaps the constants $\bic$ can be fixed so that that no $\scg^{\sn\se\sw}$ feature; 
this is termed {\it strong avoidance of integrabilities}.  

\m 

\n 2) The $\scg^{\sn\se\sw}$ are extra (Lie first-class) generators. 

\m 

\n 3) The $\scg^{\sn\se\sw}$ are specifiers. 

\m 

\n 4) The $\scg^{\sn\se\sw}$ are Lie second-class, so that brackets need redefining. 

\m 

\n 5) Perhaps the constants $\bic$ can be fixed so that that $\bTheta$ disappears;   
this is termed {\it strong avoidance of obstructions}.  

\m 

\n 6) Perhaps $\bTheta$ exceeds what can be supported by the theory. 
In this case, a `hard' obstruction is realized, killing off candidate theories rather than just modifying them.
This points to candidate theories being eliminable by topological means rather than by running out of degrees of freedom.

\section{Split Lie algebras}\label{Split}

Next suppose that a hypothesis is made about some subset of the generators $\sch$ being significant \cite{Gilmore}; 
denote the rest of the generators by $\sck$.  

\m 

\n{\bf Example 1} The $\sch$ could form a `little group' (alias stabilizer or isotropy subgroup). 

\m 

\n{\bf Example 2} The $\sch$ could form a subgroup that a contraction specifies. 

\m 

\n{\bf Example 3} The Configurational to Temporal split of Relationalism can however also be viewed in this light, 
alongside any subsequent `methodology' or `philosophy' treating linear and quadratic constraints qualitatively differently.   

\m 

\n{\bf Caveat 1} On now needs to check, however, the extent to which the algebraic structure actually complies with this assignation of significance.    
Such checks place limitations on the generality of intuitions and concepts which only hold for some simple examples of algebraic structures.

\subsection{Split at the level of a given Lie algebraic structure}\label{Split-Algebra}

\n A general split Lie algebraic structure is of the form 
\beq
\mbox{\bf |[} \,  \uc{\sch} \mbox{\bf ,} \,  \uc{\sch}\mbox{}^{\prime} \, \mbox{\bf ]|}  \es  \uc{\uc{\uc{\biA}}} \, \uc{\sch}\mbox{}^{\prime\prime}  \m + \m 
                                                                                              \uc{\uc{\uc{\biB}}} \, \uc{\sck}                          \m ,
\label{Lie-Split-1}
\eeq 
\beq
\mbox{\bf |[} \,  \uc{\sch} \mbox{\bf ,} \,  \uc{\sck}          \, \mbox{\bf ]|}  \es  \uc{\uc{\uc{\biC}}} \, \uc{\sch}\mbox{}^{\prime}               \m + \m
                                                                                       \uc{\uc{\uc{\biD}}} \, \uc{\sck}\mbox{}^{\prime}                 \m ,
\label{Lie-Split-2}
\eeq 
\beq
\mbox{\bf |[} \,  \uc{\sck} \mbox{\bf ,} \,  \uc{\sck}\mbox{}^{\prime} \, \mbox{\bf ]|}  \es  \uc{\uc{\uc{\biE}}} \, \uc{\sch}                         \m + \m               
                                                                                              \uc{\uc{\uc{\biF}}} \, \uc{\sck}\mbox{}^{\prime\prime}     \m .																	
\label{Lie-Split-3}
\eeq 
This extends \cite{ABook} e.g.\ Gilmore's split \cite{Gilmore} to include the algebroid possibility as well.  

\m 

\n{\bf Remark 1} $\biB = 0$ and $\biE = 0$ are clearly subalgebra closure conditions. 

\m 

\n{\bf Remark 2} $\biC$ and $\biD$ are `interactions between' subalgebraic structures $\Frh$ and $\Frk$, 
                          so $\biC = 0$ and $\biD = 0$ are non-interaction conditions.  

\m 

\n{\bf Remark 3} If each corresponding subalgebra condition holds, $\biA = 0$ means     that $\Frh$ is commutative, 
                                                          whereas  $\biF = 0$ signifies that $\Frk$ is commutative.  
						   
\m 						   
						   
\n The following further particular cases are realized in this Series of Articles. 

\m 

\n{\bf Structure I}   {\it Direct product} \cite{Cohn}.      
If $\biB = \biC = \biD = \bE = 0$,   then 
\be
\Frg = \Frh  \times \Frk   \m .
\ee 
\n{\bf Structure II}  {\it Semidirect product} \cite{Cohn, M63}.          
If solely $\biC \neq 0$, then 
\be 
\Frg = \Frh \rtimes \Frk   \m . 
\ee 
\n{\bf Structure III} `{\it Thomas integrability}' \cite{Gilmore}.  
If $\biB \neq 0$ is nonzero, then $\Frh$ is not a subalgebra: attempting to close it leads to some $k_{\sfk}$ are discovered to be integrabilities.
Let us denote this by 
\be 
\Frh \, \Thomas \, \Frk    \m .
\ee
A simple example of this occurs in splitting the Lorentz group's generators up into rotations and boosts: 
the group-theoretic underpinning \cite{Gilmore} of Thomas precession as per Sec VII.5.3.  

\m 

\n{\bf Structure IV}  `{\it Two-way integrability}' \cite{AMech, ABook}. 
If $\biB, \biE \neq 0$, neither $\Frh$ nor $\Frk$ are subalgebraic structures, 
due to their imposing integrabilities on each other.
Let us denote this by 
\be 
\Frh \, \TwoWay \, \Frk    \m .
\ee   
In this case, any wishes for $\Frh$ to play a significant role by itself 
are almost certainly dashed by the actual Mathematics of the algebraic structure in question.

\m 

\n{\bf Remark 1} Each step down the ladder from I) to IV) represents a large increase in complexity and generality.

\subsection{Split including integrabilities and topological terms$^*$}\label{Split-Algorithm}

\n{\bf Structure 1} We next attempt to {\sl maintain} one set of generators' $\sch$ Brackets Closure in the presence of a further disjoint set $\sck$.
We allow for new generators being discovered and topological obstruction terms arising:  
\beq
\mbox{\bf |[} \, \uc{\sch} \mbox{\bf ,} \, \uc{\sch} \, \mbox{\bf ]|}  \es  \uc{\uc{\uc{\biA}}} \, \uc{\sch}                     \m + \m 
                                                                            \uc{\uc{\uc{\biB}}} \, \uc{\sck}                     \m + \m
										    							    \uc{\uc{\uc{\biH}}} \, \uc{\sch}\mbox{}^{\sn\se\sw}  \m + \m 
																	        \uc{\uc{\uc{\biI}}} \, \uc{\sck}\mbox{}^{\sn\se\sw}  \m + \m  
																		        \uc{\uc{\bupXi}}                                   \m ,
\eeq
\beq
\mbox{\bf |[} \, \uc{\sch} \mbox{\bf ,} \, \uc{\sck} \, \mbox{\bf ]|}  \es  \uc{\uc{\uc{\biC}}} \, \uc{\sch}                     \m + \m
                                                                            \uc{\uc{\uc{\biD}}} \, \uc{\sck}                     \m + \m  
                                                                            \uc{\uc{\uc{\biJ}}} \, \uc{\sch}\mbox{}^{\sn\se\sw}  \m + \m  
	   																        \uc{\uc{\uc{\biK}}} \, \uc{\sck}\mbox{}^{\sn\se\sw}  \m + \m  
																		        \uc{\uc{\bupPhi}}                                  \m ,
\eeq
\beq
\mbox{\bf |[} \, \uc{\sck} \mbox{\bf ,} \, \uc{\sck} \, \mbox{\bf ]|}  \es  \uc{\uc{\uc{\biE}}} \, \uc{\sch}                     \m + \m  
                                                                            \uc{\uc{\uc{\biF}}} \, \uc{\sck}                     \m + \m  
																	        \uc{\uc{\uc{\biL}}} \, \uc{\sch}\mbox{}^{\sn\se\sw}  \m + \m 
																	        \uc{\uc{\uc{\biM}}} \, \uc{\sck}\mbox{}^{\sn\se\sw}  \m + \m   
																		        \uc{\uc{\bupOmega}}                                \m .
\eeq	
\n{\bf Remark 1} In all cases below, we take it without saying that further nonzero entities can be strongly removed 
as another path to each case for which these were zero in the first place.

\m 

\n{\bf Remark 2} Within this quite general ansatz, the case with no discoveries is                      
\be 
\biH  =  \biI  =  \biJ  =  \biK  =  \biL  =  \biM  =  \bupXi  =  \bupPhi  =  \bupOmega  =  0  \m ,
\ee  
\n{\bf Remark 3} The direct product case is                      
\be 
\biB  =  \biC  =  \biD  =  \biE  =  \biI  =  \biJ  =  \biK  =  \biL  =  0                       \m ,
\label{string2}
\ee 
\n{\bf Remark 4} The orientation of semidirect product which respects the $\sch$'s self-closure generalizes (\ref{string2}) further allowing for $\biD \neq 0$.  

\m 

\n{\bf Remark 5} $\biB$ or $\biI \neq 0$ means the class of $\Frh$-objects does not close as a subalgebraic structure.

\m 

\n{\bf Remark 6} $\biC \neq 0$ signifies that our algebraic structure was chosen too small for $\Frh$ to represent it.

\m 

\n Moreover, if we require that the $\sch$ represent the purported $\lFrg$ prior to bringing in the $\sck$, all of the above are moot.  

\m 

\n{\bf Remark 7} If $\biK \neq 0$, this may indicate that the $\sch$ are incompatible with the $\sck$'s $\lFrg$-invariance, 
to be resolved by the same methods as in Case 3) but now treating the $\sch$ and $\sck$ together.  

\m 

\n{\bf Remark 8}  $\biJ$ or $\biL \neq 0$ indicate that adjoining the $\sck$ to the $\sch$ forces $\lFrg$ to be extended.

\subsection{Generator Closure}\label{Closure}

The Generator Closure super-aspect now follows as per Sec III.2 and Article VII in the canonical constraints subcase, 
or Article X in the spacetime setting.
This includes a list of Generator Closure subproblems in correspondence to the diversity of outputs of the Lie Algorithm 
(or its Dirac Algorithm subcase).

\section{Assignment of Observables}\label{EitoO}

\subsection{Unrestricted observables}

\n{\bf Structure 1} The most primary notion of observables involves, given a state space $\FrS(\lFrs)$ for a system $\lFrs$, 
{\it Taking a Function Space Thereover},  $\FrF\mbox{unction}(\FrS(\lFrs))$.    

\m 

\n{\bf Structure 2} Let us first consider this in the unrestricted case, i.e.\  the absence of generators. 
The {\it unrestricted observables}  
\be
\sbiU(\bfB)                                        
\ee 
form, if working over the smooth functions, say, the {\it space of unrestricted observables}   
\be 
\UnresObs(\FrS) \es \FrC^{\infty}(\FrS(\lFrs))  \m .
\ee

\subsection{Imposition of zero commutation conditions}\label{CA}

\n{\bf Motivation 1} In a restricted theory, restricted observables are more useful for the modelling than just any function(al)s of $\bfB$, 
due to their containing between more, and solely, non-redundant modelling information.   

\m 

\n{\bf Structure 1} In the presence of generators $\scg$, the {\it commutant condition}  
\be 
\mbox{\bf |[} \, \scg  \mbox{\bf ,} \, \sbiO \, \mbox{\bf ]|}    \speq   0  
\ee
Lie brackets relation applies to the corresponding observables, $\sbiO$, indexed by $\fO$.
I.e.\
\be 
\mbox{\bf |[} \, \scg  \mbox{\bf ,} \, \sbiO \, \mbox{\bf ]|}    \es   0  
\ee
in the strong case, or 
\be  
\mbox{\bf |[} \, \uc{\scg} \mbox{\bf ,} \, \uo{\sbiO} \, \mbox{\bf ]|}  \es  \uc{\uo{\uc{\biW}}} \cdot \uc{\scg} 
\ee 
in the generator-weak case, with structure constants $\biW$, standing for `weak'. 
This undertilde is our coordinate-free notation for an index running over some type of observable.  

\m 

\n{\bf Remark 1} This forms an associated algebraic structure with respect to the same bracket operation.  

\m

\n{\bf Analogous Example 0} When the commutants are formed from the generators themselves, they are known as {\it Casimirs}.  
These play a prominent role in Representation Theory, with $SU(2)$'s total angular momentum operator $J^2$ constituting the best-known such.

\m 

\n{\bf Example 1} Geometrical observables.

\m 

\n{\bf Example 2} Canonical observables. 

\m 

\n{\bf Example 3} Spacetime observables. 

\m 

\n{\bf Lemma 1} \cite{AObs} Consistency of this commutation relation requires one's generators $\scg$ to close to form a subalgebraic structure.  

\m 

\n{\u{Proof}} This follows from the Jacobi identity 
\be
  \mbox{\bf |[} \, \sbiO  \mbox{\bf ,} \, \mbox{\bf |[} \, \scg   \mbox{\bf ,} \, \scg  \, \mbox{\bf ]|} \, \mbox{\bf ]|}   \es 
- \mbox{\bf |[} \, \scg   \mbox{\bf ,} \, \mbox{\bf |[} \, \scg   \mbox{\bf ,} \, \sbiO \, \mbox{\bf ]|} \, \mbox{\bf ]|}  
- \mbox{\bf |[} \, \scg   \mbox{\bf ,} \, \mbox{\bf |[} \, \sbiO  \mbox{\bf ,} \, \scg  \, \mbox{\bf ]|} \, \mbox{\bf ]|}  \speq  0     \m . \m  \Box
\ee  
\n{\bf Remark 1} This further Lie-algebraic relation {\sl decouples} Assignment of Observables to occur {\sl after} establishing Generator Closure.

\m 

\n{\bf Lemma 2} \cite{AObs} Observables themselves moreover close as further Lie brackets algebras.

\m 

\n{\u{Proof}}  Let $\scg$ be the generators for the defining subalgebraic structure of generators that our notion of observables $\sbiO$ corresponds to.
Then from the Jacobi identity, 
\be
  \mbox{\bf |[} \,  \scg  \mbox{\bf ,} \, \mbox{\bf |[} \, \sbiO  \mbox{\bf ,} \, \sbiO \, \mbox{\bf ]|} \, \mbox{\bf ]|}    \es 
- \mbox{\bf |[} \,  \sbiO  \mbox{\bf ,} \, \mbox{\bf |[} \, \sbiO  \mbox{\bf ,} \, \scg \, \mbox{\bf ]|} \, \mbox{\bf ]|}  
- \mbox{\bf |[} \,  \sbiO  \mbox{\bf ,} \, \mbox{\bf |[} \, \scg  \mbox{\bf ,} \, \sbiO \, \mbox{\bf ]|} \, \mbox{\bf ]|}   \speq  0     \m . \m  \Box
\ee
\n{\bf Remark 2} We write our algebra as 
\be 
\mbox{\bf|[} \, \uc{\sbiO} \mbox{\bf ,} \, \uc{\sbiO}\mbox{}^{\prime} \, \mbox{\bf ]|} = \uc{\uc{\uc{\biO}}} \, \uc{\sbiO}\mbox{}^{\prime\prime}  \m , 
\ee
for {\it observables algebra structure constants} $\biO$.

\subsection{Fully restricted observables}

\n{\bf Structure 1} The opposite extreme to imposing no restrictions is to impose all of a modelling situation's first-class generators.
This returns the {\it full observables} $\sbiF$ obeying 
%
%
\beq
\mbox{\bf |[} \, \bscF \mbox{\bf ,} \, \sbiF \, \mbox{\bf ]|}  \speq   0   \m .
\label{C-D}
\eeq 
In the canonical setting, imposing all the first-class constraints gives full observables, which are here better known as {\it Dirac observables}.  

\m 

\n{\bf Structure 2}  The {\it space of full observables} is
\be 
\FullObs(\FrS)   \m ,
\ee 
or, more explicitly, $\FullObs(\FrS(\lFrs), \cE)$. 
In the canonical case, this is alias space of Dirac observables,

\n $\DiracObs(\Phase, \cH)$.    

\m 

\n{\bf Remark 1} The unrestricted and full notions of observables are universal over all models.

\subsection{Middling observables}

First-class linear, and gauge, notions of generators continue to make sense in the general Lie case. 
See Articles VIII and X for further specifics of first-class linear and gauge observables, and spaces thereof.   

\m 

\n{\bf Remark 1} Observables algebras $\Obs$ are themselves linear spaces. 
%

\m 

\n{\bf Remark 2} Observables algebras $\Obs$, like the constraint algebraic structures, 
are comparable to configuration spaces $\FrQ$ and phase spaces $\Phase$ as regards the study of the the nature of Physical Law, 
and whose detailed structure is needed to understand any given theory.
This refers in particular to the topological, differential and higher-level geometric structures observables algebraic structures support, 
now with also function space and algebraic levels of structure relevant. 

\m 

\n{\bf Remark 3} This means we need to pay attention to the Tensor Calculus on observables algebraic structures as well, 
justifying our use of undertildes to keep observables-vectors distinct from constraints ones and spatial ones.  
In fact, the increased abstraction of each of these is `indexed' by my notation: no turns for spatial, one for constraints and two for observables.

\section{Orders and Lattices in Lie's Mathematics}\label{Order}

\n{\bf Remark 1} Applications of Order Theory have been a standard part of Algebra and Group Theory since the 30s and the 60s respectively, 
featuring e.g. in \cite{Serre-Lie} for Lie groups and Lie algebras.  
The basic structures we use are as follows. 

\m 

\n{\bf Definition 1} A {\it binary relation} $R$ on a set $\FrX$ is a property that each pair of elements of $\FrX$ may or may not possess.  
We use $a\,R\,b$ to denote `$a$ and $b$ $\in \FrX$ are related by $R$'. 

\m 

\n{\bf Remark 2} Simple examples include $=, <, \leq, \subset$ and $\subseteq$.

\m 

\n Some basic properties that an $R$ on $\FrX$ might possess are as follows ($\forall \, a, b, c \in \FrX$).  

\m 

\n{\bf Definition 2} {\it Reflexivity}:  $a\,R\,a$. 
%

\m 

\n {\it Antisymmetry}: $a\,R\,b$ and $b\,R\,a \Rightarrow a = b$.      

\m 

\n {\it Transitivity}: $a\,R\,b$ and $b\,R\,c \Rightarrow a\,R\,c$.  

\m 

\n{\it Totality}: that one or both of $a\,R\,b$ or $b\,R\,a$ holds, i.e.\ all pairs are related.  

\m 

\n{\bf Remark 3} Commonly useful combinations of these include the following.

\m


\n{\bf Definition 3} {\it Partial ordering}, $\preceq$, if $R$ is reflexive, antisymmetric and transitive. 
%

\m 

\n{\bf Definition 4} {\it Total ordering}, alias a {\it chain}, if $R$ is both a partial order and total.  

\m 

\n{\bf Definition 5} A set equipped with a partial order is termed a {\it poset} $\langle \, \FrX, \preceq \, \rangle$.  

\m 

\n{\bf Definition 6} A {\it lattice} is a poset within which each pair of elements has a least upper bound and a greatest lower bound.
In the context of a lattice, these are called {\it join} $\lor$ and {\it meet} $\land$.  

\m 

\n  An element 1 of $\lattice$ is a {\it unit}         if $\forall \m l \,  \in \,  \lattice$, $l \preceq 1$, 
and an element 0 of $\lattice$ is a {\it null element} if $\forall \m l \,  \in \,  \lattice$, $0 \preceq l$. 
A lattice that possesses these is termed a {\it bounded lattice}.
A {\it lattice morphism} is an order-, join- and meet-preserving map between lattices.  

\m 

\n{\bf Remark 4} Posets can be usefully represented by {\it Hasse diagrams}; 
all the pictures of lattices in this Series are (at least schematically) Hasse diagrams. 

\m 

\n{\bf Remark 5} See Article III for various examples of lattices of generator algebraic structures 
(some of which are constraint algebraic structures) 
and their dual lattices of observables algebraic structures.  

\m 

\n{\bf Remark 6} The sizes of the spaces run in the dual lattice pair moreover run in opposition. 
I.e.\ the bigger a constraint algebraic structure, the smaller the corresponding space of observables is. 
This is clear enough from constraints acting as restrictions, adding PDEs that the observables must satisfy.  
 
\m 
														
\n{\bf Remark 7} For some physical theories, the first-class linear constraints close algebraically, 
by which these support the corresponding notion of Kucha\v{r} observables \cite{K93}.  
For other physical theories, e.g.\ Supergravity, these do not however close by themselves, 
by which Kucha\v{r} observables are not well-defined in such theories. 
 
\m  
 
\n{\bf Remark 8} The lattice of notions of observables provides moreover a theory-independent generalization 
of the possibility of there being {\sl whatever} kinds of `middling' observables, 
a role played in GR-as-Geometrodynamics by precisely the Kucha\v{r} observables.

\section{Lie's Integral Approach to Invariants, and its uplift to an Integral Theory of Observables}\label{LIToI} 

\n{\bf Remark 1} One can obtain explicit PDEs, in the canonical case, 
by writing out what the Poisson brackets (III.65) means automatically gives first-order PDEs \cite{AObs2}. 
In the case involving generators rather than constraints, by using the derivative representation of generators has the same effect \cite{PE-1}.  
Each of these applies for all well-defined nontrivially-restricted notions of observables.   

\m 

\n{\bf Remark 2} These are first-order linear PDE systems, as detailed in Articles VIII and X.
The Flow Method is suitable for such \cite{John, Lee2}, converting such systems into coupled systems of ODEs, as per Article VIII.  
This is an outgrowth of Lagrange's Method of Characteristics\cite{Lagrange, CH2, John}; 
both this and the Flow Method itself transcend from flat space to curved manifolds $\Frm$.
The modern theory of flows is rooted on topological spaces, and differential-geometric notions such as diffeomorphisms and immersions.  
Solving flows is moreover a setting in which shrinking techniques on locality can be required.  

\m 

\n{\bf Remark 3} Unlike Sec \ref{GKEs}'s generalized Killing equations, observables equations are not affected over-determinedness 
because all the $\sbiO$ solve the zero commutation equation regardless 
($\sbiO$ is a $\bfB$-scalar and a $\lFrg$-scalar).

\m 

\n{\bf Examples} In the setting of Geometry, treating the observables PDE systems in this way 
is a slight generalization \cite{PE-1} of {\bf Lie's Integral Approach to Invariants} \cite{Lie, G63}. 
The slight generalization involves uplifting to a {\it free} alias {\it natural} \cite{CH1} characteristic problem 
for finding 'suitably-smooth functions of the invariants', i.e.\ observables.  
For canonical or spacetime physics, there is an analogous invariants problem in each case, 
and an uplifiting \cite{DO-1} to the corresponding observables problem as well.    

\m 

\n{\bf Remark 4} The free characteristic problem is appropriate since, given a state space $\bFrS$, we are Taking a Function Space Thereover: $\bFrF(\bFrS)$.  
This involves finding {\sl all} functions solving our free problem, say withing a given smoothness category such as $\FrC^{\infty}$, 
rather than a particular such corresponding to prescribed data. 

\m  

\n{\bf Remark 5} In the case of a single generator, the strong observables PDE system is just a single homogeneous-linear equation.  
The corresponding ODE system is 
\be 
\dot{x}^{\alpha} = a^{\alpha}(x^{\beta})   \m , 
\ee  
\be 
\dot{\ttO} = 0                             \m . 
\ee
Here 
\be
\dot{\m}   \:=  \frac{\d}{\d \nu}                                                                                                                       \m , 
\ee
for $\nu$ a fiducial variable to be eliminated, rather than carrying any temporal (or other geometrical or physical) significance; 
our system is moreover {\it autonomous} (none of the functions therein depend on $\nu$).  
The first block here corresponds to Lie's Integral Approach to Invariants, 
whereas coupling this to the last equation given is the strong case's uplift. 
This trivially gives that our geometrical strong observables are arbitrary suitably-smooth functions of Lie's invariants.  
Or, just as trivially, that canonical and spacetime observables are arbitrary smooth functions 
of the phase space and space-of-spacetimes analogues of Lie's invariants.  

\m 

\n{\bf Remark 5} The corresponding weak observables PDE consists of a single inhomogeneous-linear equation. 
The corresponding ODE system is now 
\be 
\dot{x}^{\alpha} = a^{\alpha}(x^{\beta}, \phi)                                                                                                        \m , 
\ee 
\be 
\dot{\sbiO} = b(x^{\beta}, \phi)                                                                                                                       \m .
\ee
From the first block being the same as before, Lie's invariants still enter our expressions for observables. 
But the inhomogeneous term means that one has further particular-integral work to do in this weak case.

\m

\n{\bf Remark 6} There is moreover a sense of middling genericity in which observables equations, 
are first-order linear {\sl systems}, 
whether homogeneous for strong observables equations, or inhomogeneous for weak observables.
There is however {\sl no} general treatment for Characteristic Problems for PDE systems of this general kind.  
Further details of the PDE system in question need to be considered in order to proceed, as follows.   

\m 

\n{\bf Remark 7} As the strong case involves a homogeneous equation, 
\be
\sbiO = \mbox{const}  \m ,
\ee
always solves; we refer to this as the {\it trivial solution}, 
and to all other solutions of first-order homogeneous quasilinear PDEs as {\it proper solutions}: 
another nontrivial kernel condition.  

\m 

\n{\bf Remark 8} Observables equation systems consist of of $G := \mbox{dim}(Aut(\Frm, \, \bsigma))$ equations for a single unknown.
So the default prima facie position is generically one of over-determination \cite{CH2}, 
leading to no solutions (or only the trivial solution, when guaranteed by homogeneity as per Remark 7).
Remarks 7 and 8 have clear generalized Killing equation counterparts.  

\m 

\n{\bf Remark 9} The way such a lack of (nontrivial) solutions might occur is via integrabilities.\footnote{While there is a conceptual counter-acting 
under-determination from the characteristicness and the freeness, it is our integrability point here that guarantees that things work out.} 
%
On the one hand, generalized Killing equations' integrability conditions \cite{Yano55} are not met generically, 
signifying that there are only any proper generalized Killing vectors at all in a zero-measure subset of $\langle \, \bFrM, \, \bsigma \, \rangle$. 
This corresponds to the generic manifold admitting no (generalized) symmetries.
On the other hand, preserved equations moreover always succeed in meeting integrability, by the following Theorem. 

\m 

\n{\bf Theorem 3} Observables equations are integrable.

\m 

\n{\bf Remark 10} This rests on an updated version of Frobenius' Theorem, as per \cite{PE-1} or Article X. 
So, while proper generalized Killing vectors generically do not exist, observables {\sl always} do, 
at least in this Series' local sense, and for sufficiently large point number $N$. 
This last caveat is clear from the examples below, 
and corresponds to zero-dimensional reduced spaces having no coordinates left to support thereover any functions of coordinates.

\m 

\n{\bf Remark 11} There is moreover a greater generality to consider: generalized Killing vector nonexistence means there are no generators to commute with.
In this case, the most primitive element of Assigning Observables -- Taking Function Spaces Thereover -- 
is manifested in a particularly simple form: taking the free functions over the state space $\FrS$. 

\m  

\n Within the secondmost-generic case -- possessing a single generalized Killing vector -- we get the single-PDE version of the problem.  

\m 

\n The $\geq 2$-compatible generalized Killing vectors case is only the next most typical, it being here that the systematic method above does not apply.

\subsection{Field-Theoretic counterpart}

These are severely disrupted by the passage from PDEs to Field Theory's {\sl functional} differential equations (FDEs).
This makes for a good mid-term focus for follow-up papers extending this Series' research; see Article VIII for further details.

\subsection{Presheaf Mathematics arises$^*$}

\n 1) Taking Function Space Thereover, and 

\m 

\n 2)  sequential restriction in solving the PDE system 

\m 

\n are evocative of presheaves \cite{Wedhorn}. It remains an interesting question whether classical observables can be modelled by the more mathematically powerful sheaves.
(Pre)sheaf theory accommodates multiple function spaces, including solution spaces of PDEs.

\subsection{Cartan's Differential Approach to Invariants}

Cartan \cite{Cartan55, G63, Olver} gave an alternative way of finding geometric invariants,  
based on differentiation and Linear Algebra within the theory of mobile frames.

\subsection{Expression in Terms of Observables}

\n {\bf Structure 1} Given a state space $\FrS$, having an observables algebraic structure $\Obs(\FrS)$ as a Function Space Thereover 
does not yet mean being able to express each model-meaningful quantity in terms of observables.  
This involves the further step of eliminating irrelevant variables in favour of observables: a matter concerning whichever of Algebra and Calculus.   
%

\m 

\n{\bf Structure 2} Spanning, independence and bases for observables algebras -- which make sense by these being linear spaces -- 
is also a significant part of the theory of observables.
It is additionally convenient to pick a basis of observables, or at least a spanning set of observables, 
in terms of which to express one's model's meaningful quantities. 
See Articles VIII, X and XI for examples. 

\m 

\n{\bf Remark 2} With \cite{AObs, ABook, PE-1, DO-1} and this Series, the days of finding individual or few observables are over. 
Solutions to the Problem of Observables are to involve, rather, 
a whole 'function space that is also an algebraic structure' of these per theory per notion of observables consistently supported by that theory. 
								
\section{Lie Algebraic Rigidity}\label{LR}

\subsection{Generator deformations}\label{LR-Def}

\n{\bf Structure 1} We denote {\it generator deformations} \cite{G64} by  
\be 
\uc{\scg} \m \longrightarrow \m \uc{\scg}\mbox{}_{\sbalpha}  \es  \uc{\scg} \m + \m  \uc{\uc{\balpha}} \cdot \uc{\bphi}  \m . 
\label{def}
\ee
$\balpha$ here in general carries a multi-index, so one has the corresponding multi-index inner product with an equally multi-indexed set of functions $\bphi$. 
These deformed generators can be viewed as terminating at linear order in $\balpha$ (rather than necessarily being small), 
since each component of our $\balpha$'s is typically a priori real-valued.  

\m 

\n{\bf Remark 1} Nijenhuis and Richardson \cite{NR66} specialized Gerstenhaber's considerations of deformations \cite{G64} to the Lie algebraic setting. 
This work additionally attributes local stability under deformations to rigid Lie algebras.

\subsection{Cohomological underpinning$^*$}\label{Cohom}

\n{\bf Structure 1} Gerstenhaber proceeds by placing a cohomological underpinning on rigidity results, 
which Nijenhuis and Richardson \cite{NR66} again specialize to the Lie algebra case as 
\be 
\mH^2(\Frg, \, \Frg) = 0  \m  \mbox{ diagnoses rigidity} \m .
\ee
This $\mH^2$ cohomology group consists of the quotient of the group 
\be 
\mZ_2(\Frg, \Frg)
\ee 
of Lie algebra 2-cocycles:  
\be 
\phi : \Frg \times \Frg \longrightarrow \Frg 
\ee
such that 
\be 
\phi(\mbox{\bf |[} \, X \mbox{\bf ,} \, Y \, \mbox{\bf ]|} , \, Z) + \mbox{cycles} - \mbox{anticycles}  \es  0
\ee 
by the group 
\be 
\mB_2(\Frg)
\ee 
of coboundaries: linear maps  
\be 
\psi: \Frg \longrightarrow \Frg 
\ee 
such that 
\be 
\phi(X, Y)  \es  (d_1 \psi)(X, Y)  \es   \psi(\mbox{\bf |[} \, X \mbox{\bf ,} \,      Y  \, \mbox{\bf ]|}) 
                                    -         \mbox{\bf |[} \, X \mbox{\bf ,} \, \psi(Y) \, \mbox{\bf ]|} 
									+         \mbox{\bf |[} \, Y \mbox{\bf ,} \, \psi(X) \, \mbox{\bf ]|}  
\ee 
for 1-coboundary $d_1$.  

\m 

\n{\bf Remark 1} The corresponding $\mH^1$ group is itself tied to the simpler matter of Lie group automorphisms.

\m 

\n{\bf Remark 2} By evoking cohomology, this treatment of deformations takes us beyond Lie's Mathematics into the terrain of Algebraic Topology 
(progressively envisaged in \cite{Poincare, Noether, DeRham, Steenrod} and reviewed in \cite{BT82, MT, Hatcher}).  
This is `global' in a further sense -- by involving the topology of some space of Lie algebraic structures -- 
thus pushing one out of this Series' main topic of A {\sl Local} Resolution of the Problem of Time.

\subsection{Discussion$^*$}\label{LR-Disc}

\n{\bf Remark 1} `Deformation' is meant here in the same kind of sense as in `deformation quantization' \cite{L78, S98, Kontsevich}.  
Such deformations are thus of some familiarity in Theoretical and Mathematical Physics.  
This Series' application is moreover a clearly distinct -- entirely classical -- application of deformation.

\m

\n{\bf Remark 2} This Series's application extends to algebroids, 
thus taking us out of Gerstenhaber's original setting \cite{G64} for deformations and rigidities -- algebras -- 
for which Nijenhuis and Richardson \cite{NR64} provided further Lie algebra specifics. 
However, e.g.\ Crainic and Moerdijk \cite{CM} subsequently considered matters of rigidity for algebroids, 
so this case remains posed and with some significant results.

\subsection{How widespread is Rigidity within the Lie Algorithm?}\label{LAR}

\n{\bf Structure 1} We next build on \cite{RWR, Phan, AM13, ABook, IX} but also \cite{G64, NR66, CM} in putting entire families of generators into the Lie Algorithm.   

\m 

\n We now furthermore identify this procedure as deformation of (some of the) input generators. 
At the level of the brackets algebraic structure itself, this sends 
\be 
\mbox{\bf |[} \, \u{\scg}\mbox{\bf ,} \, \u{\scg}^{\prime} \, \mbox{\bf ]|}  \es  \u{\u{\u{\biG}}} \, \u{\scg}^{\prime\prime}  
\ee 
to
\be   
\mbox{\bf |[} \, \u{\scg}\mbox{}_{\sbalpha}, \, \u{\scg}\mbox{}^{\prime}_{\sbalpha} \, \mbox{\bf ]|}   \es  \u{\u{\u{\biG}}}(\bfB, \balpha) \, \u{\scg}\mbox{}_{\sbalpha}^{\prime\prime}  \m + \m 
                                                                                                        \u{\u{\u{\biN}}}(\bfB, \balpha) \, \u{\scg}^{\sn\se\sw}\mbox{}_{\sbalpha}                 \m + \m  
																	                                      \u{\u{\Theta}}(\bfB, \balpha)                                               \m .
\label{alpha-Alg}
\ee
\n{\bf Remark 1} These $\u{\scg}\mbox{}^{\sn\se\sw}_{\sbalpha}$ are integrabilities of the $\u{\scg}\mbox{}_{\sbalpha}$.  

\m 

\n{\bf Remark 2} In our mathematical arena of interest, (\ref{alpha-Alg}) includes in principle the eventuality that 
a Lie algebra's structure constants $\biG$           deform to a Lie algebroid's structure functions             $\biG(\bfB , \, \balpha)$, 
or extend thereto        [i.e.\ the $\biN(\bfB , \, \balpha)$            could be structure functions even if the $\biG(\balpha)$ are not].  
The latter algebroid feature takes one out of Gerstenhaber's original algebraic setting \cite{G64}. 
Deformation thus itself gives a third reason (to Sec \ref{LAS}'s two) for involvement of Lie algebroids.  

\m 

\n{\bf Remark 3} By the final zeroth-order right-hand-side term, topological obstructions, e.g.\ along the lines of anomalies can enter proceedings.  

\m 

\n{\bf Remark 4} The $\biG(\bfB, \alpha)$, $\biN(\bfB, \alpha)$ and $\Theta(\bfB, \alpha)$ 
can moreover have strongly vanishing roots, i.e.\ particular values of $\alpha$ for which these terms disappear. 
On some occasions, this is capable of picking out special cases that remain free of topological obstructions, 
                                                                                    integrabilities, 
																					need for rebracketing, 
																					specifers, or 
																					structure functions. 

\m 

\n{\bf Remark 5}  The outcome of putting deformed Lie generators through the Lie Algorithm leaves us facing the question of when `anything goes' and when 
just one (or very few) sharp possibilities occur. 
This bears some relation to which Lie Theories are rigid under deformations. 

\m 

\n Some examples of this (Article IX) moreover also appear to realize that contracted \cite{Gilmore} limits of a given algebraic structure remain consistent, 
but have to be encoded separately from the uncontracted version.
\cite{L67} is the first known instance of contractions being treated alongside rigidities.  

\m 

\n{\bf Remark 6} We need to work with maximally general deformations to ensure that rigidity results do not disappear upon considering furtherly general deformations.

\m 

\n{\bf Remark 7} We would do well to use further examples to assess how typical is it 
for contractions of an algebraic structure to accompany uncontracted versions thereof as other means of attaining consistency in the Lie Algorithm.  

\m 

\n{\bf Remark 8} Constructability by Deformation and Rigidity splits up into, firstly, 
Space from less Space Structure Assumed and Spacetime with less Spacetime Structure Assumed, collectively referred to 
as Internal Constructabilities: internal to a given allocation of primality. 
Secondly, to Spacetime from Space, which we term {\it Primality-Transcending Constructability}.

\section{Which theories have Reallocation of Intermediary-Object Invariance?}\label{RIO}
%
{            \begin{figure}[!ht]
\centering
\includegraphics[width=0.4\textwidth]{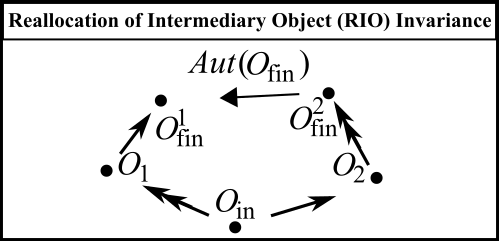}
\caption[Text der im Bilderverzeichnis auftaucht]{ \footnotesize{Commuting pentagon from $O_{\si\sn}$ to $O_{\sf\si\sn}$ via 
two distinct allocations of intermediary objects, $O_1$ and $O_2$.
The thinner arrows can be labelled by $g$ and the doubled arrows by $h$.}}
\label{Pentagon}\end{figure}            }

\n{\bf Example 1} A well-known case is Refoliation Invariance \cite{T73, K92, I93, ABook, XII} in GR [posed in Fig III.5.b) and resolved in Fig III.5.c)]. 

\m

\n This generalizes to the following universally poseable, if not necessarily realizable, structure.   

\m  

\n{\bf Definition 1} {\it Reallocation of Intermediary-Object (RIO)} is the commuting-pentagon property depicted in Figure 1.  
In more detail, it is a {\it commuting square}, corresponding to moving from an initial object $O_{\si\sn}$ to a final object $O_{\sf\si\sn}$ 
via two different intermediaries $O_{1}$ and $O_{2}$, {\it up to some automorphism of the final object}, 
\be 
Aut(O_{\sf\si\sn})  \m ,  
\ee 
relating the outcomes of proceeding via $O_{1}$ and via $O_{2}$.  
This automorphism constitutes the fifth side of the pentagon.  

\m 

\n{\bf Remark 1} One or both of $O_1$ and $O_2$ can be replaced with distinct arbitrary intermediate objects. 

\m 

\n{\bf Remark 2} Some theories will obey this property, and some will not (see Sec III.8, Article XII and \cite{Nambu} for examples). 
RIO invariance thus also has the status of a selection principle.  

\m 

\n{\bf Remark 3} We need at least one generator not among the $Aut(O_{\sf\si\sn})$, else it is trivial by $Aut$'s closure.  
In this Series, $\scC\mbox{hronos}$ plays this role.
%

\m 

\n{\bf Remark 4} RIO Invariance is a priori local, by involving the algebraic commutator rather than the group-theoretic one.
But the group-theoretic commutator can be built up from the algebraic one by Hausdorff's Lie-globalization theorem.  

\m 

\n{\bf Remark 5} Refoliation Invariance itself has the further feature of transcending back from spacetime primality to spatial primality.

\section{Conclusion}\label{Conclusion}

\subsection{Summary} 

\n We make use of a modernized Lie Theory, rooted in Topological Spaces, Toplogical Manifolds and Differentiable Manifolds, 
as is suitable for contemporary Theoretical Physics.  
This uses suitably smooth functions rather than Lie's analytic ones, and does label and elsewise quantify the 
domains and neighbourhoods in question.  
We however keep Lie's third assumption -- locality -- now extended also to state spaces, as the `local' in ALRoPoT and ALToBI.
This permits moreover transgression from differentiable manifolds to locally differentiable spaces, covering reduced state spaces;    
this is a significant robustness, insofaras ALRoPoT is a reductive approach.  
All five super-aspects of classical-level ALToBI are covered by our brand of Lie Theory, as follows.   

\m 

\n{\bf Super-aspect i)}   We set up Lie derivatives, 
which Articles I, II, III, V, VI and X employ to encode Relationalism into Principles of Dynamics actions.  
Solving the generalized Killing equation \cite{Yano55, Yano70} supports this working.   

\m  

\n{\bf Super-aspect ii)}  We consider general Lie Theory's version of Closure. 
This involves forming Lie brackets of generators provided by Relationalism, 
to be assessed by `Lie's Algorithm' (an extension of Lie's own Algorithm \cite{Lie} with insights of Dirac \cite{Dirac} and from Topology).
When successful, the output consists of Lie algebraic structures of generators, $\bFrG$.  

\m 

\n{\bf Super-aspect iii)} We next give general Lie Theory's account of observables as functions over state spaces. 
In the presence of $\bFrG$, this requires forming zero Lie brackets with $\bFrG$.  
For Finite Theories, the resulting equations can be recast as a homogeneous-linear PDE system 
to which the Flow Method is sequentially applied.
This is a slight extension of Lie's Integral Approach to Invariants.
For Field Theories, however, we get the Functional Differential Equation (FDE) counterpart, which, as Article VIII details, is a substantial research frontier.
Our solutions moreover form Lie algebras of observables.  

\m 

\n{\bf Super-aspect iv)} We also give a general theory of Constructability, 
based on deforming Lie-algebraic generators and encountering Lie-algebraic Rigidity \cite{G63, NR66}. 
This covers both internal cases -- Space from Less Structure of Space assumed, or likewise for Spacetimes -- 
and the primality-transcending case of Spacetime from Space. 

\m 

\n{\bf Super-aspect v)} We finally outline Reallocation of Intermediary Object (RIO) Invariance: 
a Lie-algebraic generalization of GR's Refoliation Invariance. 

\m 

\n This represents a major advance and simplification of the Problem of Time and Background Independence field of study.  

\m 

\n{\bf Structure 1} Each of spacetime and `space, configuration space, dynamics or canonical' primality moreover 
realizes a separate copy of Fig \ref{LSVL2}'s {\it Lie 3-star digraph}.  
This is formed from i), ii), iii) and the internal case of iv). 
Note in particular Closure's central status as the nexus alias star-point, 
whereas Relationalism can also be viewed as this directed tree's root 
(taking Relationalism as first principle and subsequently making use of it as a generator provider). 

\m 

\n A first approximate level of structure in ALRoPoT and ALToBI is thus the direct product 
\be 
\mbox{(Lie 3-star)} \times \mbox{(choice of primality)}  \m .    
\ee 
A first count for ALToBI aspects is thus $4 \times 2 = 8$.  
%
{            \begin{figure}[!ht]
\centering
\includegraphics[width=1.0\textwidth]{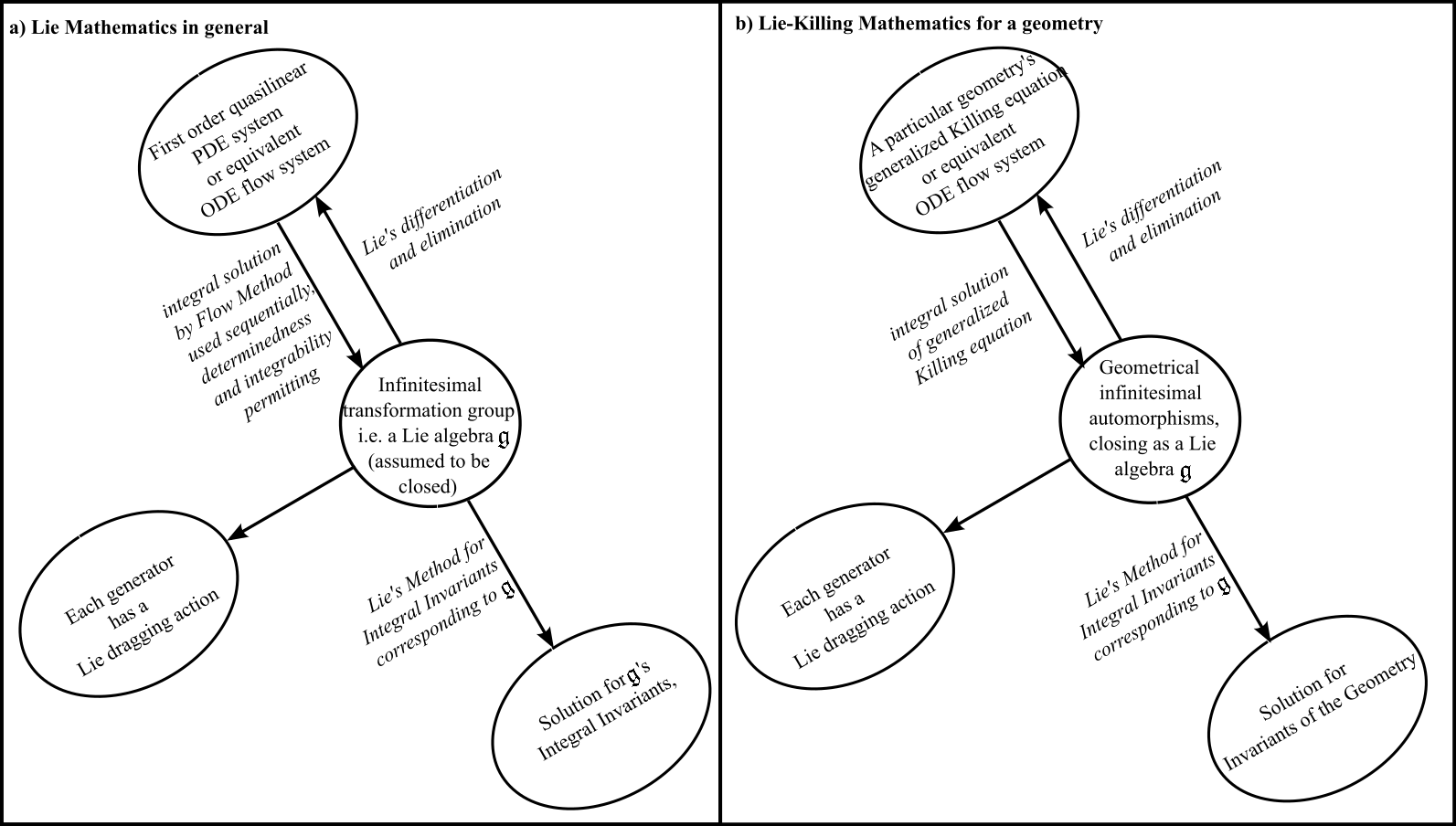}
\caption[Text der im Bilderverzeichnis auftaucht]{ \footnotesize{a) Lie's original program. 
b) Its restriction to geometries, as extended by Killing and successors \cite{Yano55, Yano70}.          }}
\label{LSVL}\end{figure}            }
%
{            \begin{figure}[!ht]
\centering
\includegraphics[width=0.7\textwidth]{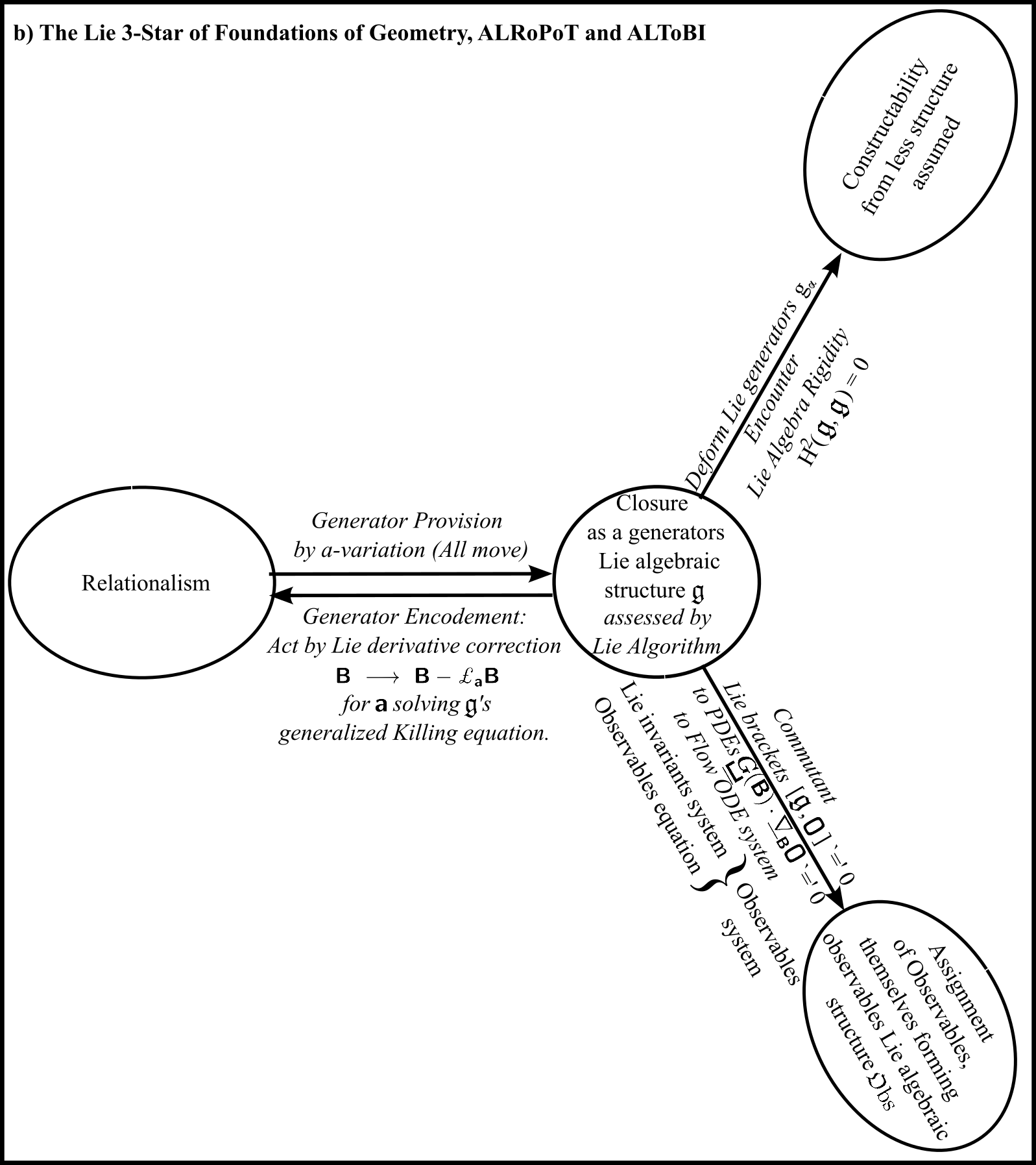}
\caption[Text der im Bilderverzeichnis auftaucht]{ \footnotesize{The Lie 3-star digraph.  
The underlying undirected graph here -- the  {\it 3-star} alias {\it claw} -- is the smallest nontrivial tree graph. 

\m 

\n The horizontal prong -- Relationalism -- concatenates b)'s leftwards prongs to form a longer prong. 
This is supported now by having an encoder function, $\bfa$-variation of which returns our Lie group.  
This will do for geometry (definite or indefinite) as well as, including a potential, spacetime Physics.
For split space-time Physics, the next figure's further modifications are necessary.  

\m 

\n The downward prong -- Assignment of Observables -- is also depicted longer, 
since   our   observables system 
extends Lie's invariants  system (but the Flow Method carries over). 

\m 

\n The upward prong -- Constructability -- postcedes Lie's and Killing's works by around 80 years as part of Lie Theory, 
by 30 further years as part of the Problem of Time \cite{RWR}, and yet another decade as a Foundation of Geometry \cite{A-Brackets}.          }}
\label{LSVL2}\end{figure}            }

\m 

\n{\bf Structure 2} Spatial primality moreover requires a separate manner of constraint production, or encodement, for Temporal Relationalism.  
This splits the horizontal prong as indicated in Fig \ref{Detail}.b).  
This has the effect of increasing ALToBI aspects from 8 to 9. 
There is now more detailed reason to loop about between Temporal and Configurational Relationalism (Fig \ref{Detail}.c).  
%
{            \begin{figure}[!ht]
\centering
\includegraphics[width=1\textwidth]{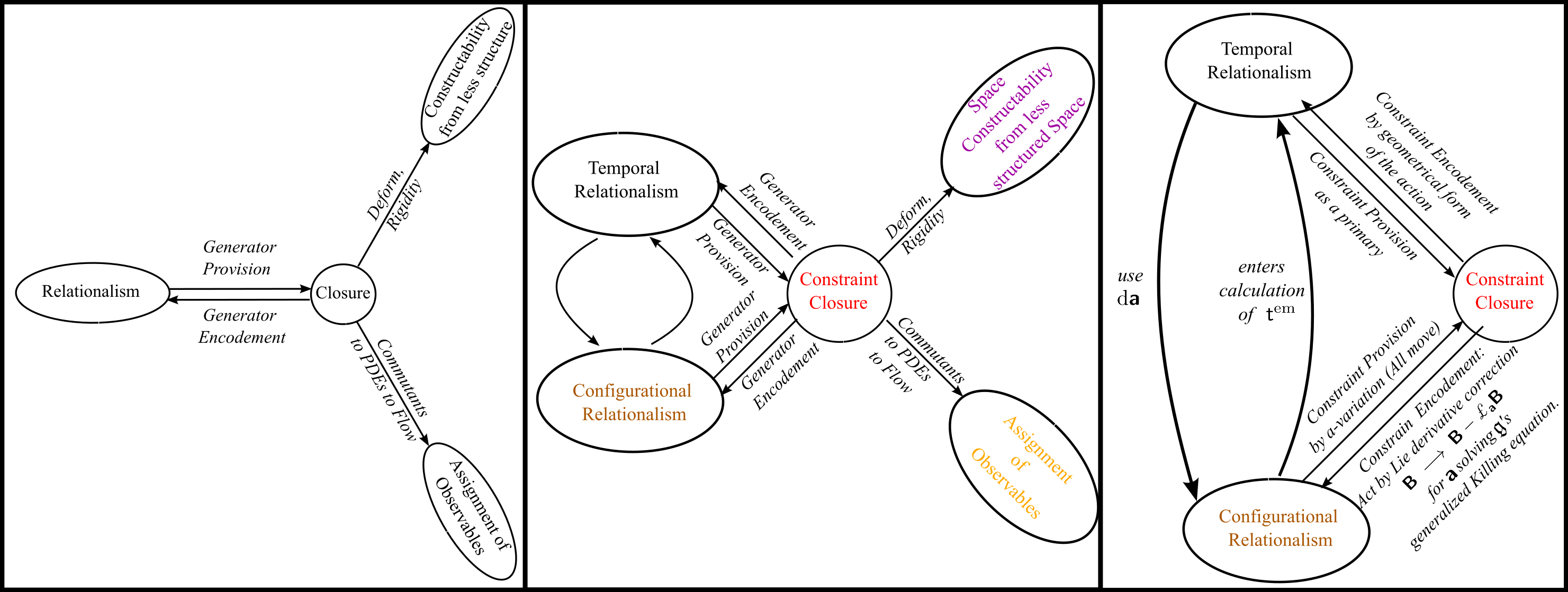}
\caption[Text der im Bilderverzeichnis auftaucht]{ \footnotesize{a) Simplified summary scheme for ALRoPoT and ALToBI's Lie 3-star digraph. 
The canonical approach's split of Relationalism in b) simplified and c) more detailed form.   }}
\label{Detail}\end{figure}            }

\m 

\n{\bf Structure 3} There is additionally a Wheelerian two-way route between the two primalities. 
This adds two aspects: Spacetime Constructability from Space, and Foliation Independence. 
These are realized by, respectively, the deformation and rigidity machinery and by Refoliation Invariance. 
The second of these is moreover supported by the ADM (\cite{ADM} and Article II) 
               [or, using compatible facets, the A  (\cite{FEPI} and Article VI)] split of the metric and action, 
as well as by `going down' by picking a foliation. 
`Going up', the Dirac Algebroid formed by the GR constraints tells us that it makes no difference which foliation was picked, 
by virtue of Refoliation Invariance.  
If GR is replaced by an arbitrary theory, RIO Invariance is the corresponding nontrivial route candidate;   
in this setting, however, neither Spacetime Constructability nor RIO Invariance are a priori guaranteed. 
These have the status, rather, of {\sl CoToBI selection principles}.    
The overall count of ALToBI aspects is thus finally adjusted to 9 + 2 = 11, 
and an answer to Kucha\v{r} and Isham's question of `in which order' to address Problem of Time facets is thus elucidated.  

\m 

\n In summary, the Problem of Time is multifaceted as a `product' of spacetime versus space primality on the one hand, 
and of {\sl Lie Theory already being multi-faceted} on the other.  
The `canonical copy' requires further elaboration due to Temporal Relationalism's distinctive nature, 
and there is also a 2-way map between primalities.

\subsection{Foundations of Geometry parallel}

\n All of super-aspects i) to v) cover Geometry, Physics with a spacetime action, and Physics in canonical form.
With this Series covering the second and third of these in detail, 
we next briefly comment on the purely-geometrical case.
\n Let us frame our account the context of Geometry arising from a number of different foundational standpoints,  
e.g.\ referred to as `Pillars of Geometry' in Stillwell's treatise \cite{Stillwell}.   

\m

\n{\bf Pillar 1)} is the Euclidean constructions approach to Geometry. 

\m 

\n{\bf Pillar 2)} is Cartesian through to Linear-Algebra-based Geometry.

\m 

\n{\bf Pillar 3)} is Klein's Erlangen Program: the transformations, groups, and invariants approach to Geometry.

\m 

\n{\bf Pillar 4)} begins with ray diagrams and leads in particular to Projective Geometry's substantial axiomatic power \cite{HC32}.  

\m

\n We furthermore view this as a list that is ever-open to additions, along similar lines to Wheeler's `many routes to relativity' \cite{Battelle, MTW}. 
On the one hand, super-aspects i) to iv) moreover point to the following additional pillars, 
some of which are well-established and others of which are new. 

\m 

\n{\bf Pillar 5)} The Lie--Killing' pillar is rather well-known \cite{Eisenhart33, Yano55}.  
This boosts the Erlangen pillar by using the generalized Killing equation to systematically construct continuous automorphisms for a given geometry.  
This furthermore parallels Configurational (or Spacetime) Relationalism: super-aspect i).  
In this regard, note the somewhat more extended form of this pillar 
in which an encoder function for the geometry is additionally entertained. 
I.e.\ a potential-less Lagrangian, or a constant-potential Jacobi arc element if Physics (or just dynamical trajectories 
through the geometry's configuration space) are to be aspired to within a fully consistent framework.  

\m 

\n{\bf Pillar 6)} is that Lie subalgebraic structures pick out \cite{A-Brackets} 
further consistently closing geometries on a given space; this parallels Closure: superaspect ii).  

\m 

\n{\bf Pillar 7)} concerns a specific flow PDE prescription for evaluating Lie's integral notion of geometric invariants, 
of which the physical Assignment of Observables -- super-aspect iii) -- is a somewhat more elaborate rendition \cite{PE-1, DO-1, VIII}. 
This `Seventh' Pillar provides zeroth principles to arrive at 
an observables version of the invariants end of the Erlangen pillar by means of solving specific PDEs: observables equations.  
This is rather reminiscent of how `continuous automorphism group finding equations' alias generalized Killing equations 
provide zeroth principles to arrive at the geometrical transformation groups end of the Erlangen program.  

\m

\n{\bf Pillar 7$^{\prime}$)} makes use instead of Cartan's Differential Approach to Invariants \cite{Cartan55}.   

\m 

\n{\bf Pillar 8)} is a Lie--Dirac pillar \cite{A-Brackets}, 
by which Dirac's inconsistency insight addendum in forming a larger Lie Algorithm for deformed generators 
confers selection principle properties that work in the presence of Lie Algebraic Rigidity: super-aspect iv).   
Article IX's Conformal versus Projective Geometry derivation of flat space's top geometry exemplifies this pillar in action. 

\m 

\n On the other hand, super-aspect v) -- RIO Invariance -- requires more than Geometry 
(or at least a `larger' notion of Geometry than just fixed finite-$d$ Geometry) to be nontrivially manifested. 
As such, it contributes a Background Independence aspect to GR, say, 
but nothing nontrivial to the Foundations of Geometry in their simplest flat-space setting.

\subsection{CoToBI$^*$} 

CoToBI appears to be mostly a global subject.  
Our outline below expands upon Epilogues II.B and III.B of \cite{ABook} on the Global Problems of Time.  
For now, CoToBI is being considered \cite{IX, XIII} in the context of differential-geometric theories with variable levels of structure, which we term `moderate' CoToBI; 
more general CoToBI involves topological Background Independence and beyond (Epilogues II.C and III.C of \cite{ABook}).    

\m 

\n Some global upgrades of basic Lie Theory machinery are as follows.
Lie groups    \cite{Serre-Lie, BCHall} in place of Lie algebras, 
global flows  \cite{John, Lee2} in place of local flows, 
with harder or more limited-applicability global DE theorems for these. 
Global fibre bundle results for Lie groups or bundles with Lie groups as whichever of structure groups or fibres \cite{Nakahara, Husemoller}.  
Some more advanced upgrades include stratification \cite{W46, T55, W65, T69, Kreck} 
and (pre)sheaves \cite{Kreck, ABook, A-Cpct} for quotient spaces by Lie groups 
-- in excess of what fibre bundle mathematics can cover -- 
and Lie groupoids \cite{Landsman} in place of Lie algebroids.  
%
%
Some particular occurrences of Global Problems of Time and Global Background Independence are as follows.  

\m 

\n{\bf Superaspect i)} Configurational and Spacetime Relationalism involve quotienting a state space $\FrS$  
by some group of physically-irrelevant geometrical transformations $Aut(\FrQ, \sigma)$ to form the {\it reduced state space}  
\be 
\w{\FrS} \:= \frac{\FrS}{Aut(\FrS, \bsigma)}     \m .   
\ee
From the point of view of general $Aut$, these are usually stratified manifolds 
(outlined in Article VI, \cite{DeWitt67, Battelle, ABook} with further details in e.g.\ 
\cite{Fischer70, FM96, ABook, A-Monopoles, PE16, A-Cpct, A-CBI}) even if $\FrS$ is itself a manifold. 
%
%
Viewed from the point of view of generic manifolds $\FrS$, however, only the trivial geometrical automorphism is supported 
-- $Aut(\FrS, \bsigma) = id$ for each $\bsigma$ -- so but a trivial quotient is realized: the manifold $\FrS$ \cite{A-Killing}. 

\m

\n When present, stratification's differential-geometrical singularness places limitations on the extent of local methods.  
This is a purely geometrical effect. 
In contrast, encoder functions $\cE$ add problems of zeros, infinities and nonsmoothnesses (Epilogue II.B in \cite{ABook}),   
with particular reference to the potential function acting as a second source for barriers to extending local treatment. 
These two limitations on extent show up side by side in GR's own Thin Sandwich \cite{BF}, 
with some indication of robustness under change of theory. 
Change of theory is however capable of compromising the elliptic nature of GR's Thin Sandwich.
Proper group actions, rooted in compactness, are moreover a major selection principle \cite{A-Cpct, A-CBI} 
in attaining  mathematically tractable quotients: LCHS (locally-compact Hausdorff second countable) stratified manifolds 
rather than merely-Kolmogorov ones.   
Fibre bundles moreover do not suffice over stratified manifolds; general bundles, presheaves, sheaves \cite{PflaumBook, Kreck} 
or differential spaces \cite{SniBook} take over this globally-significant role.  

\m  

\n{\bf Superaspect ii)} Closure's qualitative distinction as the nexus of the Lie 3-star digraph 
is strongly suggestive of CoToBI doing well to concentrate on variants of closure, at least as a first port of call. 
Aside from everything outlined in the current Article being at least {\sl poseable} 
for Supersymmetric Theories including for Supergravity in place of GR \cite{PVM, AMech, ABook}, 
we act on this by co-launching the Nijenhuis Mathematics \cite{Nijenhuis} 
                               and Nambu     Mathematics \cite{Nambu}     variants with the current Article.  
CoToBI is moroever locally influenced by each theory's particular integrability structure exhibited in closing \cite{AMech, ABook}.  

\m

\n{\bf Superaspect iii)} Global Assignment of Observables involves solving a global rather than just local PDE problem. 
This may, at least in part, be counteracted by viewing Assignment of Observables in parallel to setting up 
a sufficient number of partly-overlapping charts.
And likewise for bases of observables, so `change of basis matrices' apply on overlaps.  
(at least within setting in which just the basis goes bad but the algebraic structure of observables remains the same in both regions).   

\m

\n At the CoTOBI level, super-aspects iv) and v) remain frontiers and, moreover, selection principles. 

\m   

\n{\bf Superaspect iv)} Rigidity rests on topology, 
by introducing Gerstenhaber--Nijenhuis--Richardson \cite{G64, NR66} cohomological characterization of Lie Algebra Rigidity 
into the classical Problem of Time and Background Independence literature. 
Such cohomological characterization constitutes a further aspect of the Global Problems of Time in addition to those listed 
in Epilogues II.B and III.B of \cite{ABook}.  
Rigidity acts moreover as a selection principle: 
only some theories with candidate status of Background Independent theories would be rigid in this manner.  
CoToBI is thereby not only an algebraic venture, but more specifically a venture in Algebraic Topology.  
I.e.\ for each $\FrS$ and $\Frg$ = `$aut(\FrS, \bsigma)$ or $\bFrG(\FrS, \bsigma)$', 
which cohomology groups $\mH^2(\Frg, \Frg)$ vanish, 
leaving one with a rigid recovery of the more structured version from the less structured one?    

\m 

\n{\bf Superaspect v)} Canonical GR's possession of Refoliation Invariance is key to fleets of observers 
-- each accelerating differently and thus passing through a distinct sequence of local frames -- 
being able to reach observational concordance (Articles III and XII).  
This, or a suitable generalization, is thus a desirable feature for a Background Independent theory to possess.  
RIO Invariance is such a generalization that can be {\sl posed} for any candidate theory. 
RIO Invariance is moreover a selection principle, 
not only as regards being realized but also as regards furthermore transcending between spacetime and space primalities. 
For instance, fixed-frame or privileged-frame (sometimes alias fixed-slicing or privileged-slicing) theories do not satisfy RIO Invariance.  
How special does realizing Refoliation Invariance make GR?


\end{document}